\documentclass{article}

\usepackage[preprint]{neurips_2026}

\makeatletter
\renewcommand{\@notice}{}
\makeatother

\usepackage[utf8]{inputenc}
\usepackage[T1]{fontenc}
\usepackage{hyperref}
\usepackage{url}
\usepackage{booktabs}
\usepackage{amsfonts}
\usepackage{amsmath}
\usepackage{nicefrac}
\usepackage{microtype}
\usepackage{xcolor}
\usepackage{graphicx}
\usepackage{enumitem}
\usepackage{multirow}
\usepackage{colortbl}
\usepackage[most]{tcolorbox}
\usepackage{wrapfig}

\title{LoopCTR: Unlocking the Loop Scaling Power for Click-Through Rate Prediction}

\author{
  \textbf{Jiakai Tang$^{1,\ddagger,*}$, Runfeng Zhang$^{2,*}$, Weiqiu Wang$^{2,*}$, Yifei Liu$^{2}$, Chuan Wang$^{2}$,} \\
  \textbf{Xu Chen$^{1,\dagger}$, Yeqiu Yang$^{2,\dagger}$, Jian Wu$^{2}$, Yuning Jiang$^{2}$, Bo Zheng$^{2}$} \\
  {\normalfont $^{1}$Gaoling School of Artificial Intelligence, Renmin University of China} \\
  {\normalfont $^{2}$Alibaba Group} \\
}

\begin{document}

\maketitle
\begingroup
\renewcommand{\thefootnote}{\fnsymbol{footnote}}
\footnotetext[3]{Project leader. This work was completed during an internship at Alibaba Group.}
\footnotetext[1]{Equal contribution.}
\footnotetext[2]{Corresponding authors.}
\endgroup
\setcounter{footnote}{0}

\begin{abstract}
Scaling Transformer-based click-through rate (CTR) models by stacking more parameters brings growing computational and storage overhead, creating a widening gap between scaling ambitions and the stringent industrial deployment constraints. We propose \textbf{LoopCTR}, which introduces a \emph{loop scaling} paradigm that increases training-time computation through recursive reuse of shared model layers, decoupling computation from parameter growth. LoopCTR adopts a sandwich architecture enhanced with Hyper-Connected Residuals and Mixture-of-Experts, and employs process supervision at every loop depth to encode multi-loop benefits into the shared parameters. This enables a \emph{train-multi-loop, infer-zero-loop} strategy where a single forward pass without any loop already outperforms all baselines. Experiments on three public benchmarks and one industrial dataset demonstrate state-of-the-art performance. Oracle analysis further reveals 0.02--0.04 AUC of untapped headroom, with models trained with fewer loops exhibiting higher oracle ceilings, pointing to a promising frontier for adaptive inference.
\end{abstract}

\section{Introduction}

Inspired by the success of Transformer architectures in natural language processing, click-through rate (CTR) prediction has progressively transitioned from early deep neural network (DNN) paradigms~\citep{wang2021dcn,zhou2018deep,zhou2019deep,mao2023finalmlp} to Transformer-based frameworks~\citep{chai2025longer,dai2025onepiece,tang2025think}. Concurrently, the modeling scope has evolved from purely feature interaction modeling~\citep{song2019autoint,zhang2021multi,gui2023hiformer} to sequential user behavior modeling~\citep{xu2025climber,khrylchenko2025scaling,chen2019behavior}, and further to hybrid architectures that jointly capture feature interactions and sequential patterns~\citep{huang2026hyformer,yu2025hhft,huang2026mixformer}. This architectural evolution has established Transformers as the de facto backbone for modern CTR prediction systems.

More recently, an increasing number of industrial efforts have begun exploring scaling phenomena in the recommendation domain~\citep{zhu2025rankmixer,zhang2026zenith,jiang2026tokenmixer}, seeking to replicate the remarkable scaling laws observed in large language models (LLMs)~\citep{kaplan2020scaling,hoffmann2022training,achiam2023gpt}. Representative works such as HSTU~\citep{zhai2024actions}, MTGR~\citep{han2025mtgr}, and OneTrans~\citep{zhang2025onetrans} have investigated scaling along three principal dimensions: \emph{depth scaling} by stacking additional model layers, \emph{width scaling} by enlarging token embedding dimensions, and \emph{input scaling} by extending user historical behavior sequences to incorporate richer contextual information. These efforts consistently demonstrate a unified pattern: scaling along any of these dimensions improves downstream task performance, albeit at the cost of increased parameters, data volume, or computation.

In this work, we explore a complementary scaling dimension: \emph{computation scaling through recursive reuse}. Rather than stacking distinct parameterized layers, our core insight is to \emph{reuse the same model layers} and increase computation through \emph{recursive loop latent reasoning}. This decouples computation from parameter growth, achieving substantially better \emph{parameter efficiency} while still leveraging increased training-time computation to enhance model performance. Moreover, the recursive loop structure serves as a superior \emph{inductive bias} that mitigates overfitting on sparse recommendation data, a persistent challenge in CTR prediction.

However, realizing this vision with standard Transformer layers presents two key challenges. \textbf{First}, the static and fixed computational flow of conventional Transformer blocks limits the model's ability to iteratively refine representations across multiple loops, as the \emph{expressiveness bottleneck} constrains what a single shared layer can achieve through repeated application. \textbf{Second}, executing multiple loops at inference time still incurs proportional latency and runtime overhead, and this \emph{efficiency bottleneck} poses a significant barrier to deployment under the low-latency requirements.

To address these challenges, we propose \textbf{LoopCTR}, a simple yet effective architecture built upon a \emph{sandwich} design consisting of an Entry Block, a Loop Block, and an Exit Block, which decouple feature encoding, iterative reasoning, and score prediction, respectively. The Loop Block permits recursive input-output processing across multiple iterations. To overcome the expressiveness bottleneck, we equip the Loop Block with \emph{Hyper-Connected Residuals} and a \emph{Mixture-of-Experts (MoE)} layer that substantially expand its representational capacity. To resolve the efficiency bottleneck, we employ \emph{process supervision} that applies supervision at every loop depth, so that the multi-loop computation during training serves as a representation enhancement mechanism whose benefits are encoded into the shared parameters. At inference time, even a single forward pass without any loop produces high-quality predictions, as the model has already internalized the gains of iterative refinement.

Extensive experiments validate the effectiveness and potential of LoopCTR. We observe a clear \emph{loop scaling} effect: more loops during training consistently yield better performance, and the train-multi-loop, infer-zero-loop strategy matches or even surpasses full multi-loop inference. More notably, oracle analysis reveals \textbf{0.02--0.04 AUC} of untapped headroom\footnote{In CTR prediction, a 0.001 AUC improvement is already considered statistically and practically significant.}, and a counter-intuitive finding that models trained with \emph{fewer} loops exhibit \emph{higher} oracle ceilings. Although our current methods have not yet reached these upper bounds, this gap highlights a promising frontier for the loop architecture. Together, these results suggest that LoopCTR opens a new scaling paradigm and charts a path toward \emph{adaptive reasoning} that improves prediction quality while reducing computational cost.

In summary, our main contributions are as follows:
\begin{itemize}[leftmargin=2em]
    \item We introduce the \emph{loop scaling} paradigm for CTR prediction, which increases training-time computation through recursive reuse of shared model layers rather than stacking additional parameters, achieving a more parameter-efficient approach to scaling.
    \item We propose LoopCTR, a sandwich architecture with Hyper-Connected Residuals and MoE-augmented Loop Blocks, together with a process supervision that encodes multi-loop training benefits into shared parameters, enabling competitive performance even at zero-loop inference.
    \item We conduct comprehensive experiments demonstrating the effectiveness of LoopCTR. Furthermore, our oracle analysis reveals 0.02--0.04 AUC of untapped headroom, highlighting a promising direction for adaptive inference.
\end{itemize}

\section{Preliminary}

Given a user $u$ and a candidate item $v$, the CTR prediction task estimates the click probability $\hat{y} = p(\text{click} \mid u, v)$. The model input comprises user profile features $\mathbf{x}^{u}$ (\textit{e.g.}, user ID, age, gender, city), item features $\mathbf{x}^{v}$ (\textit{e.g.}, item ID, category, merchant, price), a short-term behavior sequence $\mathcal{S}^{s}$ capturing recent interests and a long-term behavior sequence $\mathcal{S}^{l}$ reflecting comprehensive historical preferences, context features $\mathbf{x}^{c}$ (\textit{e.g.}, device type, timestamp), and cross features $\mathbf{x}^{\times}$ encoding pre-computed user-item affinity statistics. We denote the complete input as $\mathbf{x} = (\mathbf{x}^{u}, \mathbf{x}^{v}, \mathcal{S}^{s}, \mathcal{S}^{l}, \mathbf{x}^{c}, \mathbf{x}^{\times})$.

We partition the raw features into \emph{sequential features} and \emph{global features}. Sequential features consist of the behavior sequences $\mathcal{S}^{s}$ and $\mathcal{S}^{l}$. For the short-term sequence $\mathcal{S}^{s}$, we retain the original item token representations to preserve fine-grained recent interest signals. For the long-term sequence $\mathcal{S}^{l}$, directly processing over a thousand tokens is prohibitively expensive; inspired by Q-Former~\citep{li2023blip}, we introduce a set of learnable query tokens that compress $\mathcal{S}^{l}$ into a compact representation via cross-attention, reducing downstream computational complexity while retaining salient long-term preference information. This compression is optimized end-to-end with the entire model. Global features encompass all non-sequential inputs ($\mathbf{x}^{u}$, $\mathbf{x}^{v}$, $\mathbf{x}^{c}$, $\mathbf{x}^{\times}$), each tokenized into embedding vectors. After this preprocessing, the model input can be summarized as a collection of \emph{token sequences} (short-term item tokens and long-term compressed query tokens) together with \emph{global tokens}, which we collectively denote as $\mathbf{T}$.

A CTR model $f_\theta$ maps $\mathbf{T}$ to a predicted click probability $\hat{y} = f_\theta(\mathbf{T}) \in [0, 1]$, and is optimized by minimizing the binary cross-entropy (BCE) loss over a training set $\mathcal{D} = \{(\mathbf{x}_i, y_i)\}_{i=1}^{N}$ with ground-truth labels $y_i \in \{0, 1\}$:
\begin{equation}
    \mathcal{L}_{\text{BCE}} = -\frac{1}{N}\sum_{i=1}^{N} \left[ y_i \log \hat{y}_i + (1 - y_i) \log (1 - \hat{y}_i) \right].
\end{equation}
At serving time, the model scores hundreds to thousands of candidate items per request, and the top-ranked items are selected for display to users.

\section{Methodology}

\begin{figure}[t]
    \centering
    \includegraphics[width=\linewidth]{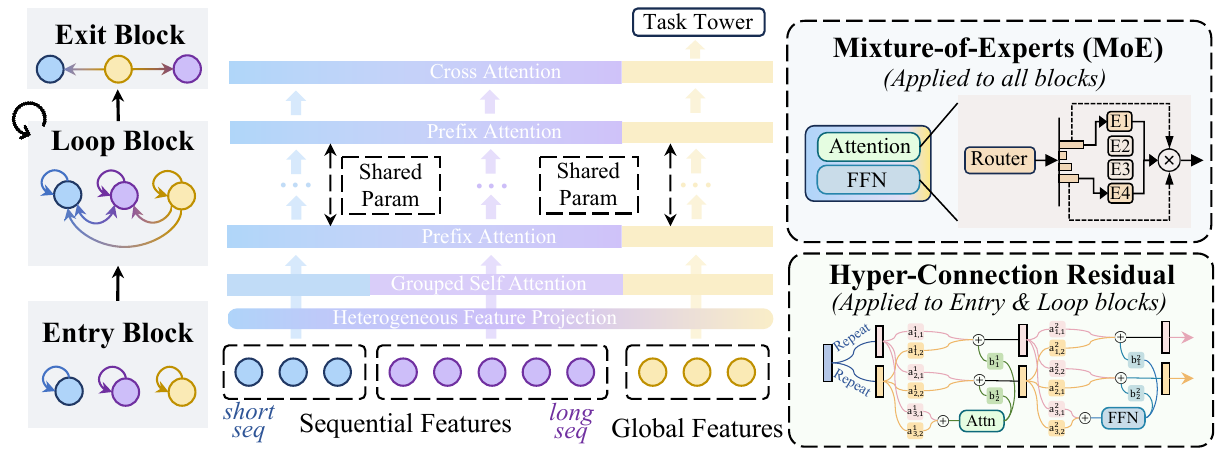}
    \vspace{-1.8em}
    \caption{Architecture of LoopCTR. \textbf{Left}: the sandwich design consisting of an \textbf{Entry Block} (heterogeneous feature projection + grouped self-attention), a \textbf{Loop Block} (prefix attention with shared parameters across iterations), and an \textbf{Exit Block} (cross-attention + task tower). \textbf{Right}: two key modules. \textit{Mixture-of-Experts (MoE)} applies sparse expert routing to both attention and FFN sub-layers across all blocks. \textit{Hyper-Connected Residuals (HCR)} provide multi-stream adaptive residual connections in the Entry and Loop Blocks, with input-dependent coefficients controlling how each stream flows through the attention and FFN sub-layers.}
    \label{fig:architecture}
\end{figure}

\subsection{Overview}

As illustrated in Figure~\ref{fig:architecture}, LoopCTR adopts a \emph{sandwich architecture} comprising three functionally distinct components: an \textbf{Entry Block} for feature encoding, a \textbf{Loop Block} for iterative latent reasoning, and an \textbf{Exit Block} for score prediction. The Entry Block encodes heterogeneous input tokens into a unified representation space. The Loop Block, which constitutes the core of our design, applies the same shared-parameter layer recursively to iteratively refine representations. The Exit Block aggregates the refined representations and produces the final click probability. This separation of concerns allows the Loop Block to be executed an arbitrary number of times during training while enabling flexible loop count reduction at inference.

\subsection{Sandwich Architecture}

\paragraph{Entry Block.}
The Entry Block performs heterogeneous feature projection followed by grouped self-attention. Since the input $\mathbf{T}$ consists of heterogeneous feature groups with distinct semantic distributions, applying a shared projection would conflate their representations. To avoid this, we employ \emph{group-specific projection matrices}: for the $g$-th token group, each token $\mathbf{t} \in \mathbb{R}^{d}$ is mapped as $\mathbf{h} = \mathbf{t} \mathbf{W}_g + \mathbf{b}_g$, where $\mathbf{W}_g \in \mathbb{R}^{d \times d'}$ and $\mathbf{b}_g \in \mathbb{R}^{d'}$ are group-specific parameters. Each behavior sequence and each individual global token constitute a separate group, so that tokens from different sources are aligned into a common feature space through distinct transformations. After this mapping, we apply full self-attention \emph{independently within} each token group. Denoting the set of token groups as $\{G_1, \dots, G_K\}$, the Entry Block output is:
\begin{equation}
    \mathbf{H}_{\text{entry}} = \big[\text{SelfAttn}(G_1); \;\dots\;; \;\text{SelfAttn}(G_K)\big],
\end{equation}
where each group is processed independently, enabling fully parallel computation.

\paragraph{Loop Block.}
The Loop Block iteratively refines token representations through $L$ repeated applications of the same shared-parameter layer. Let $\mathbf{H}_{\text{seq}}$ and $\mathbf{H}_{\text{glb}}$ denote the sequential and global token representations, respectively. At each loop iteration $l$:
\begin{equation}
    \mathbf{H}^{(l)} = \text{PrefixAttn}\!\left([\mathbf{H}_{\text{seq}}^{(l-1)}; \mathbf{H}_{\text{glb}}^{(l-1)}],\; \mathbf{M}\right),
\end{equation}
where $\text{PrefixAttn}(\cdot, \mathbf{M})$ denotes multi-head attention governed by the mask $\mathbf{M}$. The mask encodes an asymmetric attention pattern: sequential tokens attend only among themselves, while global tokens attend to the entire input. Formally, for query token $i$ and key token $j$:
\begin{equation}
    \mathbf{M}[i,j] =
    \begin{cases}
    1, & \text{if } i \in \text{seq} \text{ and } j \in \text{seq}, \\
    1, & \text{if } i \in \text{glb}, \\
    0, & \text{otherwise}.
    \end{cases}
\end{equation}
This design allows global tokens to aggregate sequential context without letting global features dominate sequential representations. To enhance the expressiveness of this single shared layer under recursive application, the Loop Block uses \emph{Hyper-Connected Residuals} (Section~\ref{sec:hyper}) and \emph{MoE-Augmented} attention and FFN (Section~\ref{sec:moe}), detailed below. In the full architecture, MoE is applied to all blocks, while HCR is applied to the Entry and Loop Blocks.

\paragraph{Exit Block.}
The Exit Block bridges iterative reasoning and final prediction. Global tokens attend to sequential tokens via cross-attention, then the global representations are concatenated and passed through an MLP:
\begin{equation}
    \hat{y} = \text{MLP}\!\left(\big[\text{CrossAttn}(\mathbf{H}_{\text{glb}}, \mathbf{H}_{\text{seq}})\big]\right).
\end{equation}

Notably, throughout the entire architecture, sequential tokens (which typically dominate the token count) never attend to global tokens (which are far fewer). This design enables KV caching for the user's sequential representations: the sequential states need only be computed once per user request and can be shared across all candidates, significantly reducing redundant computation during serving.

\subsection{Hyper-Connected Residuals}
\label{sec:hyper}

A standard Transformer block applies a fixed residual $\mathbf{h} + f(\mathbf{h})$, restricting computation to a single stream with a static 1:1 blending ratio. When the same layer is applied recursively across multiple loops, this rigid structure limits the model's ability to adaptively control information flow at different iterations. Inspired by recent advances~\citep{zhu2024hyper,xie2025mhc}, we replace the standard residual with \emph{Hyper-Connected Residuals} that extend the computation into $n$ parallel streams with input-dependent adaptive fusion.

Concretely, the single-stream hidden state $\mathbf{h} \in \mathbb{R}^{d}$ is replicated $n$ times to form a multi-stream state $\mathbf{H} \in \mathbb{R}^{n \times d}$. Given $\mathbf{H}$ and a sub-layer function $\mathcal{T}$ (\textit{e.g.}, attention or FFN), the hyper-connected update takes the form:
\begin{equation}\label{eq:hc}
    \hat{\mathbf{H}} = \underbrace{\mathbf{A}_r^{\top} \mathbf{H}}_{\text{residual mixing}} + \underbrace{\mathbf{B}^{\top} \cdot \mathcal{T}\!\left((\mathbf{H}^{\top} \mathbf{A}_m)^{\top}\right)}_{\text{layer contribution}},
\end{equation}
where $\mathbf{A}_m \in \mathbb{R}^{n \times 1}$ fuses the $n$ streams into a single input for $\mathcal{T}$, $\mathbf{B} \in \mathbb{R}^{1 \times n}$ distributes $\mathcal{T}$'s output back across streams, and $\mathbf{A}_r \in \mathbb{R}^{n \times n}$ governs the residual mixing among streams. Unlike the fixed 1:1 ratio in standard residuals, all three coefficients are \emph{input-dependent}. Let $\bar{\mathbf{H}} = \text{RMSNorm}(\mathbf{H})$, each coefficient consists of a learnable static component plus a dynamic perturbation:
\begin{equation}
    \tilde{\mathbf{A}}_m = \mathbf{A}_m + s_{\alpha} \odot \tanh(\bar{\mathbf{H}} \mathbf{W}_m), \;\;
    \tilde{\mathbf{A}}_r = \mathbf{A}_r + s_{\alpha} \odot \tanh(\bar{\mathbf{H}} \mathbf{W}_r), \;\;
    \tilde{\mathbf{B}} = \mathbf{B} + s_{\beta} \odot \tanh(\bar{\mathbf{H}} \mathbf{W}_{\beta})^{\top},
\end{equation}
where $\mathbf{A}_m$, $\mathbf{A}_r$, $\mathbf{B}$ are learnable static parameters that capture loop-invariant behavior, while the gated terms provide input-aware dynamic adjustments conditioned on the current hidden state $\bar{\mathbf{H}}$. Here $s_{\alpha}$, $s_{\beta}$ are learnable scaling factors and $\{\mathbf{W}_m, \mathbf{W}_r, \mathbf{W}_{\beta}\}$ are projection matrices. During the forward pass, the input-aware coefficients $\tilde{\mathbf{A}}_m$, $\tilde{\mathbf{A}}_r$, $\tilde{\mathbf{B}}$ substitute $\mathbf{A}_m$, $\mathbf{A}_r$, $\mathbf{B}$ in Eq.~\eqref{eq:hc}.

\paragraph{Initialization.} All projection matrices $\{\mathbf{W}_m, \mathbf{W}_r, \mathbf{W}_{\beta}\}$ are initialized to zero, so the dynamic perturbations vanish and the hyper-connection reduces to standard Pre-Norm residual at training start. The static parameters for the $s$-th sub-layer ($s=0$ for attention, $s=1$ for FFN) are set as:
\begin{equation}
    \begin{pmatrix}
    \mathbf{0}_{1\times 1} & \mathbf{B} \\
    \mathbf{A}_m & \mathbf{A}_r
    \end{pmatrix}
    =
    \begin{pmatrix}
    \mathbf{0}_{1\times 1} & \mathbf{1}_{1\times n} \\
    \mathbf{e}_{s \bmod n} & \mathbf{I}_{n\times n}
    \end{pmatrix},
\end{equation}
where $\mathbf{e}_{s \bmod n}$ is the one-hot basis vector selecting the stream assigned to the $s$-th sub-layer. Under this initialization, each sub-layer reads from exactly one designated stream and preserves all streams via the identity residual, recovering standard Pre-Norm Transformer behavior.

\subsection{MoE-Augmented Transformer}
\label{sec:moe}

While Hyper-Connected Residuals enhance the computational flow, the parameter capacity of a single shared layer may still be insufficient for capturing the diverse interaction patterns present in recommendation data. To expand the representational power without proportionally increasing computation, we integrate \emph{Mixture-of-Experts (MoE)} into attention and feed-forward components across all blocks.

\paragraph{Attention MoE.} We adopt the standard multi-head attention formulation:
\begin{equation}
    \text{Attn}(\mathbf{X}) = \text{softmax}\!\left(\frac{\mathbf{Q}\mathbf{K}^{\top}}{\sqrt{d_k}}\right) \mathbf{V} \cdot \mathbf{W}_O,
\end{equation}
where $\mathbf{Q} = \mathbf{X}\mathbf{W}_Q$, $\mathbf{K} = \mathbf{X}\mathbf{W}_K$, $\mathbf{V} = \mathbf{X}\mathbf{W}_V$, and $\mathbf{W}_O$ is the output projection. To expand the parameter capacity, we replace $\mathbf{W}_V$ and $\mathbf{W}_O$ with MoE layers, where each token is routed to a sparse subset of experts. The two MoE layers share the same router, so that each token selects the same set of experts for both value ($\mathbf{W}_V$) and output ($\mathbf{W}_O$) projections, reducing routing overhead. The Query and Key projections ($\mathbf{W}_Q$, $\mathbf{W}_K$) remain shared across all tokens to preserve consistent similarity computation.

\paragraph{FFN MoE.} Similarly, we replace the standard feed-forward network with an MoE variant, where each token is routed to its top-$k$ experts via a gating network with load-balancing auxiliary loss.

By applying MoE to both attention and FFN components, the model gains access to a substantially larger parameter pool while each token activates only a sparse subset per forward pass, maintaining computational efficiency compatible with the latency constraints of online systems. To prevent expert collapse (\textit{i.e.}, all tokens being routed to a small subset of experts), we add a load-balancing auxiliary loss that encourages uniform expert utilization (details in Appendix~\ref{app:balance_loss}).

\subsection{Training Objective}

To enable the model to produce high-quality predictions at any loop depth, we extend the standard single-point loss to a \emph{multi-depth process supervision} objective. At each loop depth $l \in \{0, 1, \dots, L\}$ (where $l=0$ corresponds to the Entry Block output before any loop iteration), the current representation is passed through the Exit Block to obtain a prediction $\hat{y}^{(l)}$. The overall training loss averages the BCE loss across all depths:
\begin{equation}
    \mathcal{L}_{\text{total}} = \frac{1}{L+1} \sum_{l=0}^{L} \mathcal{L}_{\text{BCE}}^{(l)}.
\end{equation}

\paragraph{Zero-loop inference.} Since every loop depth is explicitly supervised during training, the model learns to produce meaningful predictions even at $l=0$. At inference time, a single forward pass through the Entry and Exit Blocks alone, completely bypassing the Loop Block, yields competitive predictions while eliminating the associated latency overhead.

\paragraph{Inductive bias of weight sharing.} The shared parameters of the Loop Block must encode representations that are useful across all iteration depths, which constrains the model to learn more generalizable features. Compared to stacking heterogeneous layers with distinct parameters, this weight-sharing structure reduces overfitting on sparse recommendation data while achieving equivalent computational depth through iterative latent reasoning.

\section{Experiments}

In this section, we conduct extensive experiments to answer the following research questions:
\textbf{RQ1}: How does LoopCTR perform compared to existing CTR prediction methods?
\textbf{RQ2}: How do training and inference loop counts affect performance?
\textbf{RQ3}: What is the contribution of each core component in LoopCTR?
Additional analyses including per-loop training diagnostics, MoE parameter sensitivity, and expert routing behavior are provided in the appendix.

\subsection{Experimental Setup}

\paragraph{Datasets.}
In line with prior work, we evaluate on three widely-used public CTR benchmarks: \textbf{Amazon (Electronics)}~\citep{he2016ups}, \textbf{TaobaoAds}, and \textbf{KuaiVideo}~\citep{10.1145/3343031.3350950}. We additionally construct an \textbf{InHouse} dataset sampled from nine days of production logs (2026/01/21--2026/01/29) at a leading e-commerce platform, which uniquely includes long-term user behavior sequences (up to 1024). Dataset statistics are summarized in Table~\ref{tab:datasets}.

\paragraph{Baselines.}
We compare LoopCTR against three categories of baselines: (1)~\textbf{DNN-based methods}: DLRM~\citep{10.1145/2959100.2959190}, DIN~\citep{zhou2018deep}, DCNv2~\citep{wang2021dcn}, and Wukong~\citep{zhang2024wukong}; (2)~\textbf{Transformer-based feature interaction}: DHEN~\citep{zhang2022dhen}, AutoInt~\citep{song2019autoint}, and HiFormer~\citep{gui2023hiformer}; (3)~\textbf{Unified sequence and feature modeling}: InterFormer~\citep{10.1145/3746252.3761527}, OneTrans~\citep{zhang2025onetrans}, HSTU~\citep{zhai2024actions}, and MTGR~\citep{han2025mtgr}. We also include \textbf{StackCTR}, a variant that replaces the shared-parameter Loop Block with 3 heterogeneous layers (each iteration uses distinct parameters), matched to LoopCTR(3/3) in FLOPs (iso-FLOPs comparison), serving as a direct comparison between loop-based parameter reuse and conventional layer stacking.

\paragraph{Evaluation Metrics.}
We adopt three standard CTR evaluation metrics: \textbf{AUC} (Area Under the ROC Curve), \textbf{GAUC} (Group AUC, computed per-user and averaged), and \textbf{NE} (Normalized Entropy), defined as the average log-loss normalized by the entropy of the empirical CTR distribution.

\begin{table}[t]
\centering
\caption{Statistics of the four evaluation datasets. Seq.\ / Non-Seq.\ denotes the number of sequential and non-sequential feature fields, respectively. Max Seq.\ Len.\ reports the maximum behavior sequence length; InHouse includes both short-term (50) and long-term (1024) sequences.}
\label{tab:datasets}
\begin{tabular}{l r r r r}
\toprule
\textbf{Statistic} & \textbf{Amazon} & \textbf{TaobaoAds} & \textbf{KuaiVideo} & \textbf{InHouse} \\
\midrule
\# Interactions  & 2,993,570  & 25,029,426  & 13,661,383  & 6,115,949  \\
\# Users         & 63,001     & 1,061,768   & 10,000      & 600,000    \\
\# Items         & 192,403    & 827,006     & 3,240,282   & 2,081,279  \\
\# Fields (Seq.\ / Non-Seq.) & 2 / 4 & 3 / 19 & 4 / 5 & 8 / 33 \\
Max Seq.\ Len.   & 100        & 50          & 100         & 50 / 1024  \\
\bottomrule
\end{tabular}
\end{table}

\definecolor{loopbg}{RGB}{232,243,255}
\definecolor{oraclebg}{RGB}{255,255,230}
\definecolor{L0bg}{RGB}{245,245,245}
\definecolor{L1bg}{RGB}{232,243,255}
\definecolor{L2bg}{RGB}{255,237,225}
\definecolor{L3bg}{RGB}{230,248,230}

\subsection{Overall Performance (RQ1)}

Table~\ref{tab:overall} presents the prediction quality of all methods across four datasets; the corresponding efficiency comparison (parameters, FLOPs, latency) is provided in Appendix Table~\ref{tab:efficiency}. We highlight the \textbf{best} and \underline{second-best} results (excluding Oracle). LoopCTR($i$/$L$) denotes inference with $i$ loops out of $L$ training loops. We make the following observations.

\begin{table}[t]
\centering
\caption{Overall performance comparison. \textbf{Bold}: best; \underline{underline}: second-best (excluding Oracle). LoopCTR($i$/$L$): $i$ inference loops / $L$ training loops. $\uparrow$: higher is better; $\downarrow$: lower is better. All improvements over the best baseline are statistically significant with $p < 0.05$ under a paired $t$-test.}
\label{tab:overall}
\resizebox{\linewidth}{!}{
\setlength{\tabcolsep}{3pt}
\begin{tabular}{l|ccc|ccc|ccc|ccc}
\toprule
\multirow{2}{*}{\textbf{Method}} & \multicolumn{3}{c|}{\textbf{Amazon}} & \multicolumn{3}{c|}{\textbf{TaobaoAds}} & \multicolumn{3}{c|}{\textbf{KuaiVideo}} & \multicolumn{3}{c}{\textbf{InHouse}} \\
\cmidrule(lr){2-4} \cmidrule(lr){5-7} \cmidrule(lr){8-10} \cmidrule(lr){11-13}
& AUC\,$\uparrow$ & GAUC\,$\uparrow$ & NE\,$\downarrow$ & AUC\,$\uparrow$ & GAUC\,$\uparrow$ & NE\,$\downarrow$ & AUC\,$\uparrow$ & GAUC\,$\uparrow$ & NE\,$\downarrow$ & AUC\,$\uparrow$ & GAUC\,$\uparrow$ & NE\,$\downarrow$ \\
\midrule
DLRM        & .8537 & .8516 & .6937 & .6388 & .5629 & .9765 & .7360 & .6449 & .8858 & .6971 & .5673 & .9226 \\
DIN         & .8587 & .8415 & .8234 & .6386 & .5626 & .9743 & .7430 & .6604 & .8811 & .7000 & .5653 & .9208 \\
DCNv2       & .8611 & .8592 & .6804 & .6408 & .5663 & .9721 & .7418 & .6565 & .8836 & .6989 & .5674 & .9209 \\
Wukong      & .8663 & .8647 & .6612 & .6388 & .5658 & .9774 & .7426 & .6629 & .8913 & .6945 & .5639 & .9236 \\
\midrule
DHEN        & .8600 & .8582 & .6776 & .6392 & .5654 & .9749 & .7375 & .6504 & .8886 & .6977 & .5678 & .9225 \\
AutoInt     & .8529 & .8517 & .6874 & .6430 & .5637 & .9843 & .7385 & .6512 & .8819 & .6965 & .5661 & .9248 \\
HiFormer    & .8615 & .8595 & .6876 & .6273 & .5608 & .9894 & .7336 & .6440 & .9003 & .6925 & .5587 & .9275 \\
\midrule
OneTrans    & .8689 & .8670 & .6678 & .6412 & .5659 & .9746 & .7417 & .6578 & .8786 & .6985 & .5649 & .9230 \\
HSTU        & .8657 & .8639 & .6712 & .6352 & .5609 & .9736 & .7373 & .6482 & .8850 & .6960 & .5445 & .9271 \\
InterFormer & .8386 & .8373 & .7187 & .6352 & .5645 & .9782 & .7245 & .6257 & .8989 & .6874 & .5516 & .9274 \\
MTGR        & .8554 & .8544 & .6859 & .6370 & .5643 & .9729 & .7349 & .6440 & .8871 & .6984 & .5662 & .9210 \\
\midrule
StackCTR(3) & .8690 & .8679 & .6605 & .6403 & \textbf{.5674} & .9732 & .7423 & .6624 & .8843 & .6999 & .5681 & .9194 \\
\rowcolor{L3bg}
LoopCTR(0/3) & .8715 & .8700 & .6594 & .6439 & \underline{.5666} & .9694 & .7448 & .6635 & \textbf{.8774} & \textbf{.7007} & .5687 & \underline{.9187} \\
\rowcolor{L3bg}
LoopCTR(1/3) & \textbf{.8728} & \underline{.8713} & .6571 & \textbf{.6441} & .5664 & \textbf{.9691} & \textbf{.7450} & \textbf{.6640} & \textbf{.8774} & \textbf{.7007} & \textbf{.5691} & \textbf{.9185} \\
\rowcolor{L3bg}
LoopCTR(2/3) & \textbf{.8728} & \textbf{.8715} & \underline{.6567} & \underline{.6440} & .5664 & \textbf{.9691} & \underline{.7449} & \underline{.6639} & \textbf{.8774} & \underline{.7005} & \underline{.5690} & \underline{.9187} \\
\rowcolor{L3bg}
LoopCTR(3/3) & \underline{.8726} & \underline{.8713} & \textbf{.6560} & .6436 & .5662 & \underline{.9693} & .7448 & .6638 & \textbf{.8774} & .7002 & .5688 & .9190 \\
\midrule
\rowcolor{oraclebg}
\textit{Oracle} & \textit{.8858} & \textit{.8844} & \textit{.6207} & \textit{.6672} & \textit{.6303} & \textit{.9569} & \textit{.7588} & \textit{.6942} & \textit{.8591} & \textit{.7195} & \textit{.6414} & \textit{.9013} \\
\bottomrule
\end{tabular}
}
\end{table}

\paragraph{LoopCTR establishes new state-of-the-art across all benchmarks.}
As shown in Table~\ref{tab:overall}, LoopCTR variants sweep the top ranks on AUC and NE across all four datasets, outperforming both traditional DNN methods and recent Transformer-based approaches. The gains are particularly notable against the strongest baselines in each category: on Amazon, LoopCTR(1/3) surpasses OneTrans by 0.0039 AUC (0.8728 vs.\ 0.8689); on KuaiVideo, it exceeds DIN by 0.0020 AUC (0.7450 vs.\ 0.7430). These improvements are meaningful by CTR standards, where even a 0.001 AUC gain carries practical significance. Importantly, the advantage holds across datasets of varying scale and domain, from the 3M-interaction Amazon dataset to the 25M-interaction TaobaoAds, suggesting that the loop scaling paradigm captures a generalizable learning principle.

\paragraph{Zero-loop inference already outperforms all baselines.}
A striking finding is that LoopCTR(0/3), which bypasses the Loop Block entirely at inference, already surpasses all baselines on AUC and NE across every dataset. On InHouse, LoopCTR(0/3) achieves the best AUC with only 13.38M FLOPs and 9.26ms latency, while HSTU requires 2150M FLOPs / 775.72ms and OneTrans requires 417.97M FLOPs / 494.58ms (Appendix Table~\ref{tab:efficiency}). The gap between zero-loop and the best multi-loop variant is marginal (0.0013 AUC on Amazon), confirming that process supervision successfully encodes the benefits of iterative refinement into the shared parameters during training.

\paragraph{Shared parameters generalize better than stacked layers.}
Comparing LoopCTR(3/3) with StackCTR under identical FLOPs reveals a clear advantage for weight sharing: LoopCTR(3/3) leads on AUC across all four datasets (0.8726 vs.\ 0.8690 on Amazon; 0.7002 vs.\ 0.6999 on InHouse). Although StackCTR occasionally matches or exceeds LoopCTR (\textit{e.g.}, TaobaoAds), the overall pattern indicates that shared parameters serve as a stronger inductive bias, forcing the model to learn representations that generalize across loop depths rather than overfitting at a fixed depth.

\paragraph{Oracle analysis reveals an order-of-magnitude headroom.}
Post-hoc oracle selection of the optimal loop depth per sample uncovers 0.013--0.023 AUC of untapped performance over the best realized LoopCTR($i/3$) result across datasets. On TaobaoAds, the oracle achieves 0.6672 AUC, a 0.0231 gap above the best realized result (0.6441). This headroom represents a substantial frontier that future adaptive inference strategies could exploit.

\subsection{Loop Scaling (RQ2)}
\label{sec:loop_scaling}

To understand how training and inference loop counts interact, we sweep training loops $L \in \{0, 1, 2, 3\}$ and, for each $L$, evaluate all inference loop counts $i \in \{0, 1, 2, 3\}$. Figure~\ref{fig:loop_scaling} visualizes the AUC trends; full results are in Appendix Table~\ref{tab:loop_scaling_full}.

\begin{figure}[t]
    \centering
    \includegraphics[width=\linewidth]{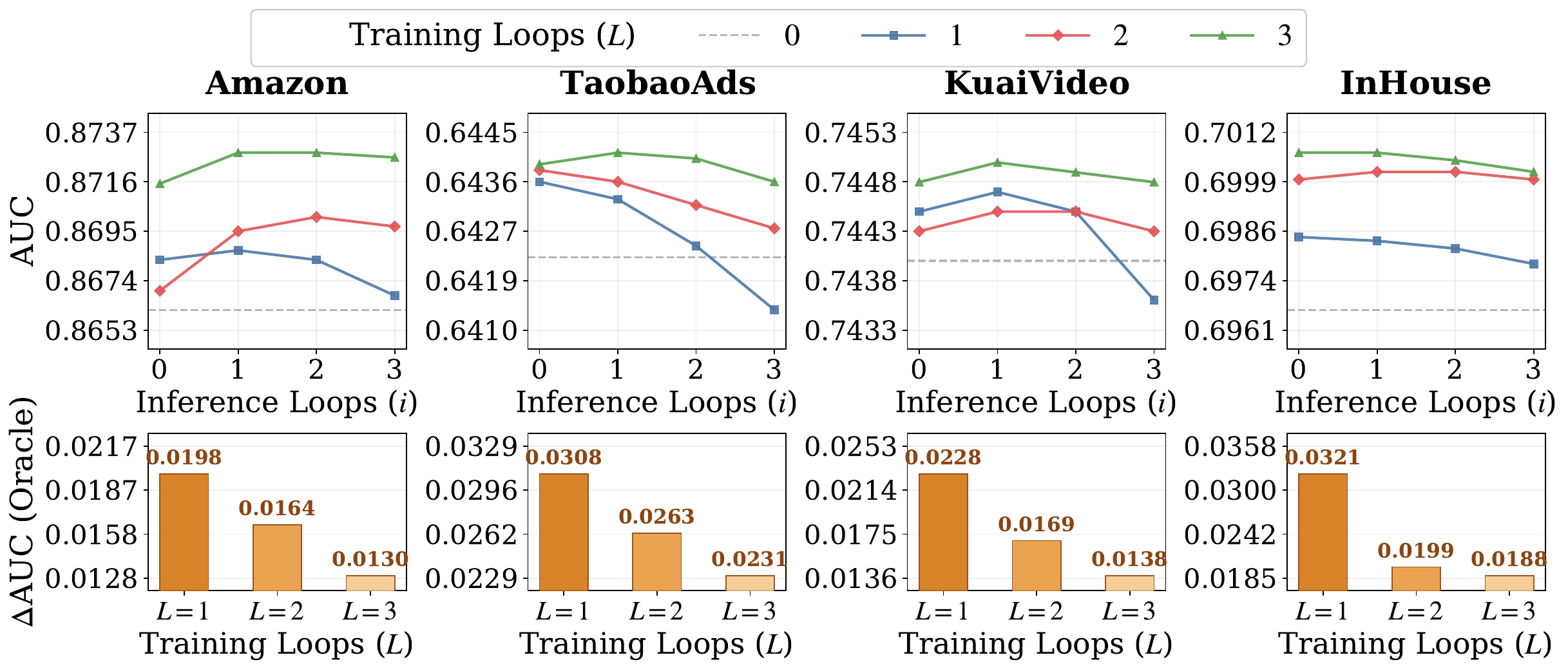}
    \vspace{-1em}
    \caption{Loop scaling analysis across four datasets. \textbf{Top row}: AUC under different training loop counts $L$ (colored lines) and inference loop counts $i$ (x-axis). The gray dashed line marks the $L{=}0$ baseline. \textbf{Bottom row}: Oracle headroom ($\Delta$AUC between oracle and best realized inference) at each $L$. All oracle results are computed with $i{=}3$ inference loops; for $L{<}3$ this constitutes extrapolation beyond the training loop count. Fewer training loops yield higher oracle ceilings.}
    \label{fig:loop_scaling}
\end{figure}

\paragraph{More training loops improve realized performance.}
Increasing $L$ from 0 to 3 consistently raises AUC across all datasets (\textit{e.g.}, 0.8662$\to$0.8728 on Amazon, 0.6966$\to$0.7007 on InHouse), confirming that deeper loop scaling consistently improves model quality. Oracle analysis further reveals 0.02--0.04 AUC of total headroom above the $L{=}0$ baseline, of which current realized gains capture only a fraction. The loss landscape visualization (Figure~\ref{fig:loss_landscape} in Appendix~\ref{app:loss_landscape}) shows that more training loops produce broader, flatter minima, explaining the improved generalization.

\paragraph{Inference loops exhibit diminishing returns.}
Within each training configuration, the first inference loop provides the largest improvement, while additional loops yield marginal or no further gains. On Amazon with $L{=}3$, AUC rises from 0.8715 ($i{=}0$) to 0.8728 ($i{=}1$) but plateaus at 0.8728 ($i{=}2$) and slightly dips to 0.8726 ($i{=}3$). This pattern holds across all datasets, reinforcing the viability of zero-loop or single-loop inference for practical deployment.

\paragraph{Fewer training loops yield higher oracle ceilings.}
A counter-intuitive but consistent finding emerges: the oracle performance ceiling \emph{increases} as $L$ decreases. On Amazon, the oracle AUC is 0.8858 at $L{=}3$, 0.8865 at $L{=}2$, and 0.8885 at $L{=}1$. On InHouse, the pattern is even more pronounced: 0.7195 ($L{=}3$) vs.\ 0.7306 ($L{=}1$). The loss landscape visualization (Figure~\ref{fig:loss_landscape} in Appendix~\ref{app:loss_landscape}) offers a geometric explanation: while models trained with more loops achieve flatter, more generalizable minima (explaining their higher realized performance), this flatness also homogenizes representations across loop depths. In contrast, models trained with fewer loops develop sharper, more concentrated minima, but with greater representational diversity across depths, providing richer opportunities for per-sample adaptive loop selection. Although realizing this oracle potential remains an open challenge, the finding underscores the substantial headroom within the loop architecture.

\subsection{Ablation Study (RQ3)}

\begin{figure}[t]
    \centering
    \includegraphics[width=\linewidth]{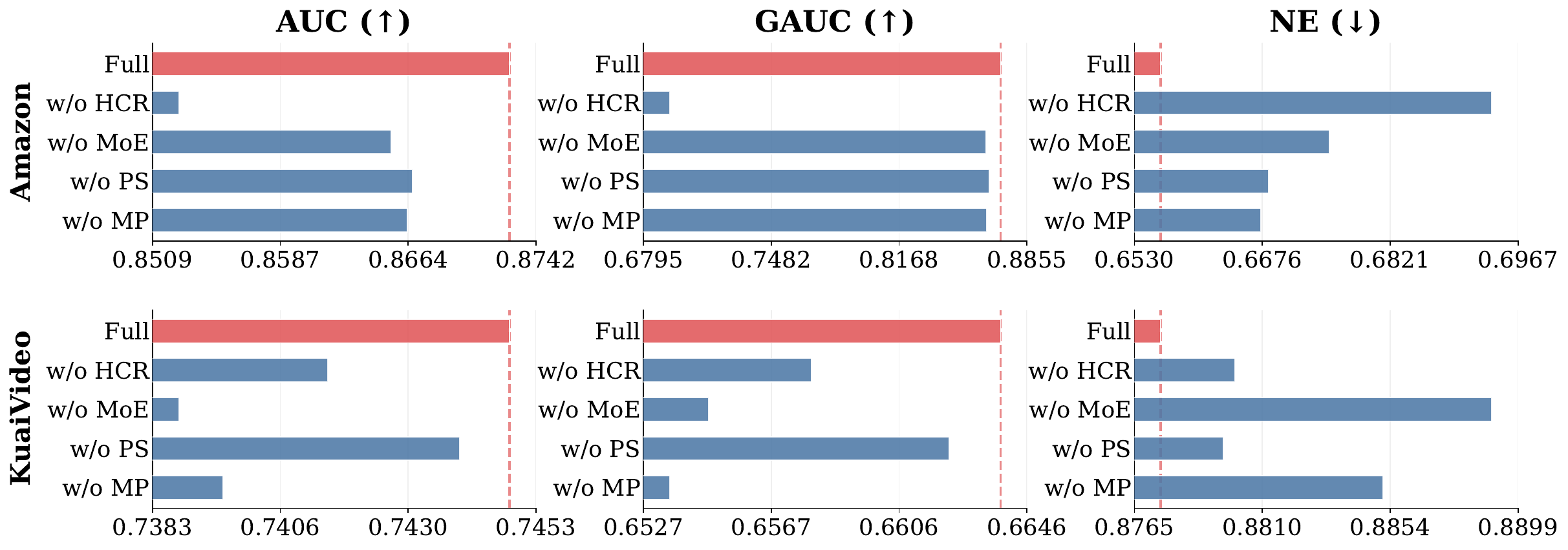}
    \vspace{-1em}
    \caption{Ablation study on Amazon (top) and KuaiVideo (bottom). Each variant removes one component from the full LoopCTR(3/3) model (red bar). The red dashed line marks the full model performance. HCR: Hyper-Connected Residuals; MoE: Mixture-of-Experts; PS: process supervision; MP: heterogeneous feature projection in the Entry Block.}
    \label{fig:ablation}
\end{figure}

To evaluate the contribution of each core component, we conduct ablation experiments on Amazon and KuaiVideo by removing one component at a time from the full LoopCTR(3/3) model. As shown in Figure~\ref{fig:ablation}, all four components contribute positively, though their relative importance varies across datasets. On Amazon, Hyper-Connected Residuals are the most critical component (removing them drops AUC by 0.0201), while on KuaiVideo, MoE has the largest impact (AUC drops by 0.0060). This suggests that the adaptive residual flow is essential for effective recursive computation, whereas the expanded parameter capacity from MoE becomes more important for datasets with richer sequential patterns. Process supervision and heterogeneous token projection contribute consistently on both datasets, validating that each component addresses a distinct bottleneck in the loop scaling paradigm.

\section{Related Work}

\paragraph{Transformer-based CTR Prediction.}
CTR prediction has evolved from feature interaction networks~\citep{zhou2018deep,wang2021dcn,mao2023finalmlp} through self-attention models~\citep{song2019autoint,chen2019behavior,gui2023hiformer} to industrial-scale Transformer systems that pursue scaling along depth, width, and input length~\citep{zhai2024actions,zhang2025onetrans,huang2026hyformer,huang2026mixformer}. While these approaches have achieved strong results, they uniformly couple additional parameters with additional computation. LoopCTR instead decouples the two by recursively reusing shared layers, scaling computation without growing the parameter footprint.

\paragraph{Looped Transformers.}
The Universal Transformer~\citep{dehghani2018universal} first proposed recursively applying shared-weight blocks with Adaptive Computation Time~\citep{graves2016adaptive}. Recent work has established theoretical foundations~\citep{xu2024expressive,saunshi2025reasoning} and demonstrated practical benefits for length generalization~\citep{fan2024looped}. These efforts originate from the NLP community and focus on language modeling or algorithmic reasoning tasks. However, they all require executing multiple loops at inference time, incurring proportional latency overhead that is prohibitive for latency-sensitive applications such as recommender systems. LoopCTR brings the loop scaling concept into the CTR prediction domain with a tailored architecture (sandwich design with Hyper-Connected Residuals and MoE) and a process supervision strategy that together enable a train-multi-loop, infer-zero-loop paradigm, addressing both the expressiveness limitation of naive weight sharing and the inference cost barrier, thereby opening a new scaling dimension for CTR prediction. A  more comprehensive discussion is provided in Appendix~\ref{app:related_work}.

\section{Conclusion}

We present LoopCTR, which introduces a loop scaling paradigm for CTR prediction. By recursively reusing shared model layers, LoopCTR decouples computation scaling from parameter growth, offering a fundamentally different path from the prevailing approach of stacking more parameters. The sandwich architecture, combined with Hyper-Connected Residuals, Mixture-of-Experts, and process supervision, enables a train-multi-loop, infer-zero-loop strategy that achieves state-of-the-art prediction quality with substantially lower inference cost. Extensive experiments validate the effectiveness and practical benefits of LoopCTR.
Our oracle analysis reveals a significant gap between realized and optimal per-sample loop selection, suggesting that the loop architecture harbors considerable untapped potential. We believe adaptive inference strategies that dynamically allocate loop depth per sample represent a promising direction for future work. Additionally, system-level optimizations such as FlashAttention and mixed-precision training/inference can be readily integrated to further improve both training and inference efficiency.

\bibliographystyle{plainnat}
\bibliography{references}

\begin{thebibliography}{51}
\providecommand{\natexlab}[1]{#1}
\providecommand{\url}[1]{\texttt{#1}}
\expandafter\ifx\csname urlstyle\endcsname\relax
  \providecommand{\doi}[1]{doi: #1}\else
  \providecommand{\doi}{doi: \begingroup \urlstyle{rm}\Url}\fi

\bibitem[Achiam et~al.(2023)Achiam, Adler, Agarwal, Ahmad, Akkaya, Aleman, Almeida, Altenschmidt, Altman, Anadkat, et~al.]{achiam2023gpt}
Josh Achiam, Steven Adler, Sandhini Agarwal, Lama Ahmad, Ilge Akkaya, Florencia~Leoni Aleman, Diogo Almeida, Janko Altenschmidt, Sam Altman, Shyamal Anadkat, et~al.
\newblock Gpt-4 technical report.
\newblock \emph{arXiv preprint arXiv:2303.08774}, 2023.

\bibitem[Chai et~al.(2025)Chai, Ren, Xiao, Yang, Han, Zhang, Chen, Lu, Zhao, Yu, et~al.]{chai2025longer}
Zheng Chai, Qin Ren, Xijun Xiao, Huizhi Yang, Bo~Han, Sijun Zhang, Di~Chen, Hui Lu, Wenlin Zhao, Lele Yu, et~al.
\newblock Longer: Scaling up long sequence modeling in industrial recommenders.
\newblock In \emph{Proceedings of the Nineteenth ACM Conference on Recommender Systems}, pages 247--256, 2025.

\bibitem[Chen et~al.(2019)Chen, Zhao, Li, Huang, and Ou]{chen2019behavior}
Qiwei Chen, Huan Zhao, Wei Li, Pipei Huang, and Wenwu Ou.
\newblock Behavior sequence transformer for e-commerce recommendation in alibaba.
\newblock In \emph{Proceedings of the 1st international workshop on deep learning practice for high-dimensional sparse data}, pages 1--4, 2019.

\bibitem[Covington et~al.(2016)Covington, Adams, and Sargin]{10.1145/2959100.2959190}
Paul Covington, Jay Adams, and Emre Sargin.
\newblock Deep neural networks for youtube recommendations.
\newblock In \emph{Proceedings of the 10th ACM Conference on Recommender Systems}, RecSys '16, page 191–198, New York, NY, USA, 2016. Association for Computing Machinery.
\newblock ISBN 9781450340359.
\newblock \doi{10.1145/2959100.2959190}.
\newblock URL \url{https://doi.org/10.1145/2959100.2959190}.

\bibitem[Csord{\'a}s et~al.(2024)Csord{\'a}s, Irie, Schmidhuber, Potts, and Manning]{csordas2024moeut}
R{\'o}bert Csord{\'a}s, Kazuki Irie, J{\"u}rgen Schmidhuber, Christopher Potts, and Christopher~D Manning.
\newblock Moeut: Mixture-of-experts universal transformers.
\newblock \emph{Advances in Neural Information Processing Systems}, 37:\penalty0 28589--28614, 2024.

\bibitem[Dai et~al.(2025)Dai, Tang, Wu, Wang, Zhu, Chen, Hong, Zhao, Fu, Wu, et~al.]{dai2025onepiece}
Sunhao Dai, Jiakai Tang, Jiahua Wu, Kun Wang, Yuxuan Zhu, Bingjun Chen, Bangyang Hong, Yu~Zhao, Cong Fu, Kangle Wu, et~al.
\newblock Onepiece: Bringing context engineering and reasoning to industrial cascade ranking system.
\newblock \emph{arXiv preprint arXiv:2509.18091}, 2025.

\bibitem[Dehghani et~al.(2018)Dehghani, Gouws, Vinyals, Uszkoreit, and Kaiser]{dehghani2018universal}
Mostafa Dehghani, Stephan Gouws, Oriol Vinyals, Jakob Uszkoreit, and {\L}ukasz Kaiser.
\newblock Universal transformers.
\newblock \emph{arXiv preprint arXiv:1807.03819}, 2018.

\bibitem[Fan et~al.(2024)Fan, Du, Ramchandran, and Lee]{fan2024looped}
Ying Fan, Yilun Du, Kannan Ramchandran, and Kangwook Lee.
\newblock Looped transformers for length generalization.
\newblock \emph{arXiv preprint arXiv:2409.15647}, 2024.

\bibitem[Fedus et~al.(2022)Fedus, Zoph, and Shazeer]{fedus2022switch}
William Fedus, Barret Zoph, and Noam Shazeer.
\newblock Switch transformers: Scaling to trillion parameter models with simple and efficient sparsity.
\newblock \emph{Journal of Machine Learning Research}, 23\penalty0 (120):\penalty0 1--39, 2022.

\bibitem[Giannou et~al.(2023)Giannou, Rajput, Sohn, Lee, Lee, and Papailiopoulos]{giannou2023looped}
Angeliki Giannou, Shashank Rajput, Jy-yong Sohn, Kangwook Lee, Jason~D Lee, and Dimitris Papailiopoulos.
\newblock Looped transformers as programmable computers.
\newblock In \emph{International Conference on Machine Learning}, pages 11398--11442. PMLR, 2023.

\bibitem[Graves(2016)]{graves2016adaptive}
Alex Graves.
\newblock Adaptive computation time for recurrent neural networks.
\newblock \emph{arXiv preprint arXiv:1603.08983}, 2016.

\bibitem[Gui et~al.(2023)Gui, Wang, Yin, Jin, Kula, Xu, Hong, and Chi]{gui2023hiformer}
Huan Gui, Ruoxi Wang, Ke~Yin, Long Jin, Maciej Kula, Taibai Xu, Lichan Hong, and Ed~H Chi.
\newblock Hiformer: Heterogeneous feature interactions learning with transformers for recommender systems.
\newblock \emph{arXiv preprint arXiv:2311.05884}, 2023.

\bibitem[Han et~al.(2025)Han, Yin, Chen, Jiang, Jiang, Li, Ma, Huang, Li, Jing, et~al.]{han2025mtgr}
Ruidong Han, Bin Yin, Shangyu Chen, He~Jiang, Fei Jiang, Xiang Li, Chi Ma, Mincong Huang, Xiaoguang Li, Chunzhen Jing, et~al.
\newblock Mtgr: Industrial-scale generative recommendation framework in meituan.
\newblock In \emph{Proceedings of the 34th ACM International Conference on Information and Knowledge Management}, pages 5731--5738, 2025.

\bibitem[He and McAuley(2016)]{he2016ups}
Ruining He and Julian McAuley.
\newblock Ups and downs: Modeling the visual evolution of fashion trends with one-class collaborative filtering.
\newblock In \emph{proceedings of the 25th international conference on world wide web}, pages 507--517, 2016.

\bibitem[Hoffmann et~al.(2022)Hoffmann, Borgeaud, Mensch, Buchatskaya, Cai, Rutherford, Casas, Hendricks, Welbl, Clark, et~al.]{hoffmann2022training}
Jordan Hoffmann, Sebastian Borgeaud, Arthur Mensch, Elena Buchatskaya, Trevor Cai, Eliza Rutherford, DDL Casas, Lisa~Anne Hendricks, Johannes Welbl, Aidan Clark, et~al.
\newblock Training compute-optimal large language models.
\newblock \emph{arXiv preprint arXiv:2203.15556}, 10, 2022.

\bibitem[Huang et~al.(2026{\natexlab{a}})Huang, Zhang, Fan, Huang, Wei, Chai, Ni, Zheng, and Chen]{huang2026mixformer}
Xu~Huang, Hao Zhang, Zhifang Fan, Yunwen Huang, Zhuoxing Wei, Zheng Chai, Jinan Ni, Yuchao Zheng, and Qiwei Chen.
\newblock Mixformer: Co-scaling up dense and sequence in industrial recommenders.
\newblock \emph{arXiv preprint arXiv:2602.14110}, 2026{\natexlab{a}}.

\bibitem[Huang et~al.(2026{\natexlab{b}})Huang, Hong, Xiao, Jin, Luo, Wang, Chai, Wu, Zheng, and Lin]{huang2026hyformer}
Yunwen Huang, Shiyong Hong, Xijun Xiao, Jinqiu Jin, Xuanyuan Luo, Zhe Wang, Zheng Chai, Shikang Wu, Yuchao Zheng, and Jingjian Lin.
\newblock Hyformer: Revisiting the roles of sequence modeling and feature interaction in ctr prediction.
\newblock \emph{arXiv preprint arXiv:2601.12681}, 2026{\natexlab{b}}.

\bibitem[Jiang et~al.(2026)Jiang, Zhu, Han, Lu, Bai, Yang, Wu, Zhang, Zhao, Bai, et~al.]{jiang2026tokenmixer}
Yuchen Jiang, Jie Zhu, Xintian Han, Hui Lu, Kunmin Bai, Mingyu Yang, Shikang Wu, Ruihao Zhang, Wenlin Zhao, Shipeng Bai, et~al.
\newblock Tokenmixer-large: Scaling up large ranking models in industrial recommenders.
\newblock \emph{arXiv preprint arXiv:2602.06563}, 2026.

\bibitem[Kaplan et~al.(2020)Kaplan, McCandlish, Henighan, Brown, Chess, Child, Gray, Radford, Wu, and Amodei]{kaplan2020scaling}
Jared Kaplan, Sam McCandlish, Tom Henighan, Tom~B Brown, Benjamin Chess, Rewon Child, Scott Gray, Alec Radford, Jeffrey Wu, and Dario Amodei.
\newblock Scaling laws for neural language models.
\newblock \emph{arXiv preprint arXiv:2001.08361}, 2020.

\bibitem[Khrylchenko et~al.(2025)Khrylchenko, Matveev, Makeev, and Baikalov]{khrylchenko2025scaling}
Kirill Khrylchenko, Artem Matveev, Sergei Makeev, and Vladimir Baikalov.
\newblock Scaling recommender transformers to one billion parameters.
\newblock \emph{arXiv preprint arXiv:2507.15994}, 2025.

\bibitem[Koishekenov et~al.(2025)Koishekenov, Lipani, and Cancedda]{koishekenov2025encode}
Yeskendir Koishekenov, Aldo Lipani, and Nicola Cancedda.
\newblock Encode, think, decode: Scaling test-time reasoning with recursive latent thoughts.
\newblock \emph{arXiv preprint arXiv:2510.07358}, 2025.

\bibitem[Lai et~al.(2023)Lai, Chen, Yeh, Xu, Cai, and Yang]{10.1145/3604915.3608831}
Vivian Lai, Huiyuan Chen, Chin-Chia~Michael Yeh, Minghua Xu, Yiwei Cai, and Hao Yang.
\newblock Enhancing transformers without self-supervised learning: A loss landscape perspective in sequential recommendation.
\newblock In \emph{Proceedings of the 17th ACM Conference on Recommender Systems}, RecSys '23, page 791–797, New York, NY, USA, 2023. Association for Computing Machinery.
\newblock ISBN 9798400702419.
\newblock \doi{10.1145/3604915.3608831}.
\newblock URL \url{https://doi.org/10.1145/3604915.3608831}.

\bibitem[Lee et~al.(2024)Lee, Willette, Kim, and Hwang]{lee2024visualizinglosslandscapeselfsupervised}
Youngwan Lee, Jeffrey~Ryan Willette, Jonghee Kim, and Sung~Ju Hwang.
\newblock Visualizing the loss landscape of self-supervised vision transformer, 2024.
\newblock URL \url{https://arxiv.org/abs/2405.18042}.

\bibitem[Li et~al.(2018)Li, Xu, Taylor, Studer, and Goldstein]{li2018visualizing}
Hao Li, Zheng Xu, Gavin Taylor, Christoph Studer, and Tom Goldstein.
\newblock Visualizing the loss landscape of neural nets.
\newblock \emph{Advances in neural information processing systems}, 31, 2018.

\bibitem[Li et~al.(2023)Li, Li, Savarese, and Hoi]{li2023blip}
Junnan Li, Dongxu Li, Silvio Savarese, and Steven Hoi.
\newblock Blip-2: Bootstrapping language-image pre-training with frozen image encoders and large language models.
\newblock In \emph{International conference on machine learning}, pages 19730--19742. PMLR, 2023.

\bibitem[Li et~al.(2019)Li, Liu, Yin, Cui, Xu, and Nie]{10.1145/3343031.3350950}
Yongqi Li, Meng Liu, Jianhua Yin, Chaoran Cui, Xin-Shun Xu, and Liqiang Nie.
\newblock Routing micro-videos via a temporal graph-guided recommendation system.
\newblock In \emph{Proceedings of the 27th ACM International Conference on Multimedia}, MM '19. Association for Computing Machinery, 2019.
\newblock ISBN 9781450368896.
\newblock \doi{10.1145/3343031.3350950}.
\newblock URL \url{https://doi.org/10.1145/3343031.3350950}.

\bibitem[Loshchilov and Hutter(2017)]{loshchilov2017decoupled}
Ilya Loshchilov and Frank Hutter.
\newblock Decoupled weight decay regularization.
\newblock \emph{arXiv preprint arXiv:1711.05101}, 2017.

\bibitem[Mao et~al.(2023)Mao, Zhu, Su, Cai, Li, and Dong]{mao2023finalmlp}
Kelong Mao, Jieming Zhu, Liangcai Su, Guohao Cai, Yuru Li, and Zhenhua Dong.
\newblock Finalmlp: an enhanced two-stream mlp model for ctr prediction.
\newblock In \emph{Proceedings of the AAAI conference on artificial intelligence}, volume~37, pages 4552--4560, 2023.

\bibitem[Na et~al.(2022)Na, Mehta, and Strubell]{na-etal-2022-train}
Clara Na, Sanket~Vaibhav Mehta, and Emma Strubell.
\newblock Train flat, then compress: Sharpness-aware minimization learns more compressible models.
\newblock In Yoav Goldberg, Zornitsa Kozareva, and Yue Zhang, editors, \emph{Findings of the Association for Computational Linguistics: EMNLP 2022}, pages 4909--4936, Abu Dhabi, United Arab Emirates, December 2022. Association for Computational Linguistics.
\newblock \doi{10.18653/v1/2022.findings-emnlp.361}.

\bibitem[Saunshi et~al.(2025)Saunshi, Dikkala, Li, Kumar, and Reddi]{saunshi2025reasoning}
Nikunj Saunshi, Nishanth Dikkala, Zhiyuan Li, Sanjiv Kumar, and Sashank~J Reddi.
\newblock Reasoning with latent thoughts: On the power of looped transformers.
\newblock \emph{arXiv preprint arXiv:2502.17416}, 2025.

\bibitem[Song et~al.(2019)Song, Shi, Xiao, Duan, Xu, Zhang, and Tang]{song2019autoint}
Weiping Song, Chence Shi, Zhiping Xiao, Zhijian Duan, Yewen Xu, Ming Zhang, and Jian Tang.
\newblock Autoint: Automatic feature interaction learning via self-attentive neural networks.
\newblock In \emph{Proceedings of the 28th ACM international conference on information and knowledge management}, pages 1161--1170, 2019.

\bibitem[Sui et~al.(2025)Sui, Chuang, Wang, Zhang, Zhang, Yuan, Liu, Wen, Zhong, Zou, et~al.]{sui2025stop}
Yang Sui, Yu-Neng Chuang, Guanchu Wang, Jiamu Zhang, Tianyi Zhang, Jiayi Yuan, Hongyi Liu, Andrew Wen, Shaochen Zhong, Na~Zou, et~al.
\newblock Stop overthinking: A survey on efficient reasoning for large language models.
\newblock \emph{arXiv preprint arXiv:2503.16419}, 2025.

\bibitem[Tang et~al.(2025)Tang, Dai, Shi, Xu, Chen, Chen, Wu, and Jiang]{tang2025think}
Jiakai Tang, Sunhao Dai, Teng Shi, Jun Xu, Xu~Chen, Wen Chen, Jian Wu, and Yuning Jiang.
\newblock Think before recommend: Unleashing the latent reasoning power for sequential recommendation.
\newblock \emph{arXiv preprint arXiv:2503.22675}, 2025.

\bibitem[Wang et~al.(2021)Wang, Shivanna, Cheng, Jain, Lin, Hong, and Chi]{wang2021dcn}
Ruoxi Wang, Rakesh Shivanna, Derek Cheng, Sagar Jain, Dong Lin, Lichan Hong, and Ed~Chi.
\newblock Dcn v2: Improved deep \& cross network and practical lessons for web-scale learning to rank systems.
\newblock In \emph{Proceedings of the web conference 2021}, pages 1785--1797, 2021.

\bibitem[Xie et~al.(2025)Xie, Wei, Cao, Zhao, Deng, Li, Dai, Gao, Chang, Yu, et~al.]{xie2025mhc}
Zhenda Xie, Yixuan Wei, Huanqi Cao, Chenggang Zhao, Chengqi Deng, Jiashi Li, Damai Dai, Huazuo Gao, Jiang Chang, Kuai Yu, et~al.
\newblock mhc: Manifold-constrained hyper-connections.
\newblock \emph{arXiv preprint arXiv:2512.24880}, 2025.

\bibitem[Xu and Sato(2024)]{xu2024expressive}
Kevin Xu and Issei Sato.
\newblock On expressive power of looped transformers: Theoretical analysis and enhancement via timestep encoding.
\newblock \emph{arXiv preprint arXiv:2410.01405}, 2024.

\bibitem[Xu et~al.(2025)Xu, Wang, Guo, Guo, Xiao, Huang, Wu, and Luo]{xu2025climber}
Songpei Xu, Shijia Wang, Da~Guo, Xianwen Guo, Qiang Xiao, Bin Huang, Guanlin Wu, and Chuanjiang Luo.
\newblock Climber: Toward efficient scaling laws for large recommendation models.
\newblock In \emph{Proceedings of the 34th ACM International Conference on Information and Knowledge Management}, pages 6193--6200, 2025.

\bibitem[Yu et~al.(2025)Yu, Zhang, Zhou, Zhang, Zhang, and Ou]{yu2025hhft}
Liren Yu, Wenming Zhang, Silu Zhou, Tao Zhang, Zhixuan Zhang, and Dan Ou.
\newblock Hhft: Hierarchical heterogeneous feature transformer for recommendation systems.
\newblock \emph{arXiv preprint arXiv:2511.20235}, 2025.

\bibitem[Zeng et~al.(2025)Zeng, Liu, Hang, Liu, Zhou, Yang, Liu, Ruan, Chen, Chen, Hao, Xu, Nie, Liu, Zhang, Wen, Yuan, Yin, Zhang, Wang, Chen, Han, Li, Yang, Long, Yu, Tong, and Yang]{10.1145/3746252.3761527}
Zhichen Zeng, Xiaolong Liu, Mengyue Hang, Xiaoyi Liu, Qinghai Zhou, Chaofei Yang, Yiqun Liu, Yichen Ruan, Laming Chen, Yuxin Chen, Yujia Hao, Jiaqi Xu, Jade Nie, Xi~Liu, Buyun Zhang, Wei Wen, Siyang Yuan, Hang Yin, Xin Zhang, Kai Wang, Wen-Yen Chen, Yiping Han, Huayu Li, Chunzhi Yang, Bo~Long, Philip~S. Yu, Hanghang Tong, and Jiyan Yang.
\newblock Interformer: Effective heterogeneous interaction learning for click-through rate prediction.
\newblock In \emph{Proceedings of the 34th ACM International Conference on Information and Knowledge Management}, CIKM '25, page 6225–6233. Association for Computing Machinery, 2025.
\newblock ISBN 9798400720406.
\newblock \doi{10.1145/3746252.3761527}.
\newblock URL \url{https://doi.org/10.1145/3746252.3761527}.

\bibitem[Zhai et~al.(2024)Zhai, Liao, Liu, Wang, Li, Cao, Gao, Gong, Gu, He, et~al.]{zhai2024actions}
Jiaqi Zhai, Lucy Liao, Xing Liu, Yueming Wang, Rui Li, Xuan Cao, Leon Gao, Zhaojie Gong, Fangda Gu, Michael He, et~al.
\newblock Actions speak louder than words: Trillion-parameter sequential transducers for generative recommendations.
\newblock \emph{arXiv preprint arXiv:2402.17152}, 2024.

\bibitem[Zhang et~al.(2022)Zhang, Luo, Liu, Li, Chen, Zhang, Wei, Hao, Tsang, Wang, et~al.]{zhang2022dhen}
Buyun Zhang, Liang Luo, Xi~Liu, Jay Li, Zeliang Chen, Weilin Zhang, Xiaohan Wei, Yuchen Hao, Michael Tsang, Wenjun Wang, et~al.
\newblock Dhen: A deep and hierarchical ensemble network for large-scale click-through rate prediction.
\newblock \emph{arXiv preprint arXiv:2203.11014}, 2022.

\bibitem[Zhang et~al.(2024)Zhang, Luo, Chen, Nie, Liu, Guo, Zhao, Li, Hao, Yao, et~al.]{zhang2024wukong}
Buyun Zhang, Liang Luo, Yuxin Chen, Jade Nie, Xi~Liu, Daifeng Guo, Yanli Zhao, Shen Li, Yuchen Hao, Yantao Yao, et~al.
\newblock Wukong: Towards a scaling law for large-scale recommendation.
\newblock \emph{arXiv preprint arXiv:2403.02545}, 2024.

\bibitem[Zhang et~al.(2021)Zhang, Qian, Cui, Liu, Li, Zhou, Ma, and Chen]{zhang2021multi}
Kai Zhang, Hao Qian, Qing Cui, Qi~Liu, Longfei Li, Jun Zhou, Jianhui Ma, and Enhong Chen.
\newblock Multi-interactive attention network for fine-grained feature learning in ctr prediction.
\newblock In \emph{Proceedings of the 14th ACM international conference on web search and data mining}, pages 984--992, 2021.

\bibitem[Zhang et~al.(2026)Zhang, Huang, Wang, Sun, Zheng, Jiang, Chen, Ouyang, Xie, Shen, et~al.]{zhang2026zenith}
Ruifeng Zhang, Zexi Huang, Zikai Wang, Ke~Sun, Bohang Zheng, Yuchen Jiang, Zhe Chen, Zhen Ouyang, Huimin Xie, Phil Shen, et~al.
\newblock Zenith: Scaling up ranking models for billion-scale livestreaming recommendation.
\newblock \emph{arXiv preprint arXiv:2601.21285}, 2026.

\bibitem[Zhang et~al.(2023)Zhang, Xue, Zhang, Zhang, Wang, Cheng, Song, and Song]{10205050}
Tianli Zhang, Mengqi Xue, Jiangtao Zhang, Haofei Zhang, Yu~Wang, Lechao Cheng, Jie Song, and Mingli Song.
\newblock Generalization matters: Loss minima flattening via parameter hybridization for efficient online knowledge distillation.
\newblock In \emph{2023 IEEE/CVF Conference on Computer Vision and Pattern Recognition (CVPR)}, pages 20176--20185, 2023.
\newblock \doi{10.1109/CVPR52729.2023.01932}.

\bibitem[Zhang et~al.(2025)Zhang, Pei, Guo, Wang, Feng, Sun, Liu, and Sun]{zhang2025onetrans}
Zhaoqi Zhang, Haolei Pei, Jun Guo, Tianyu Wang, Yufei Feng, Hui Sun, Shaowei Liu, and Aixin Sun.
\newblock Onetrans: Unified feature interaction and sequence modeling with one transformer in industrial recommender.
\newblock \emph{arXiv preprint arXiv:2510.26104}, 2025.

\bibitem[Zhou et~al.(2018)Zhou, Zhu, Song, Fan, Zhu, Ma, Yan, Jin, Li, and Gai]{zhou2018deep}
Guorui Zhou, Xiaoqiang Zhu, Chenru Song, Ying Fan, Han Zhu, Xiao Ma, Yanghui Yan, Junqi Jin, Han Li, and Kun Gai.
\newblock Deep interest network for click-through rate prediction.
\newblock In \emph{Proceedings of the 24th ACM SIGKDD international conference on knowledge discovery \& data mining}, pages 1059--1068, 2018.

\bibitem[Zhou et~al.(2019)Zhou, Mou, Fan, Pi, Bian, Zhou, Zhu, and Gai]{zhou2019deep}
Guorui Zhou, Na~Mou, Ying Fan, Qi~Pi, Weijie Bian, Chang Zhou, Xiaoqiang Zhu, and Kun Gai.
\newblock Deep interest evolution network for click-through rate prediction.
\newblock In \emph{Proceedings of the AAAI conference on artificial intelligence}, volume~33, pages 5941--5948, 2019.

\bibitem[Zhu et~al.(2024)Zhu, Huang, Huang, Zeng, Mao, Wu, Min, and Zhou]{zhu2024hyper}
Defa Zhu, Hongzhi Huang, Zihao Huang, Yutao Zeng, Yunyao Mao, Banggu Wu, Qiyang Min, and Xun Zhou.
\newblock Hyper-connections.
\newblock \emph{arXiv preprint arXiv:2409.19606}, 2024.

\bibitem[Zhu et~al.(2025{\natexlab{a}})Zhu, Fan, Zhu, Jiang, Wang, Han, Ding, Wang, Zhao, Gong, et~al.]{zhu2025rankmixer}
Jie Zhu, Zhifang Fan, Xiaoxie Zhu, Yuchen Jiang, Hangyu Wang, Xintian Han, Haoran Ding, Xinmin Wang, Wenlin Zhao, Zhen Gong, et~al.
\newblock Rankmixer: Scaling up ranking models in industrial recommenders.
\newblock In \emph{Proceedings of the 34th ACM International Conference on Information and Knowledge Management}, pages 6309--6316, 2025{\natexlab{a}}.

\bibitem[Zhu et~al.(2025{\natexlab{b}})Zhu, Wang, Hua, Zhang, Li, Que, Wei, Wen, Yin, Xing, et~al.]{zhu2025scaling}
Rui-Jie Zhu, Zixuan Wang, Kai Hua, Tianyu Zhang, Ziniu Li, Haoran Que, Boyi Wei, Zixin Wen, Fan Yin, He~Xing, et~al.
\newblock Scaling latent reasoning via looped language models.
\newblock \emph{arXiv preprint arXiv:2510.25741}, 2025{\natexlab{b}}.

\end{thebibliography}

\newpage

\appendix

\begin{tcolorbox}[
    enhanced,
    colback=white,
    colframe=black!60,
    boxrule=0.5pt,
    arc=0pt,
    top=8pt, bottom=8pt, left=12pt, right=12pt,
    before skip=6pt, after skip=10pt
]
\centerline{\textbf{Appendix Table of Contents}}
\vspace{6pt}
\hrule height 0.3pt
\vspace{6pt}
\begin{itemize}[leftmargin=1.5em, itemsep=5pt, parsep=0pt, topsep=2pt, label={}]
    \item \textbf{Appendix~\ref{app:loop_deep_dive}.\; Loop Scaling Deep Dive}
    \begin{itemize}[leftmargin=1.2em, itemsep=0pt, parsep=0pt, topsep=1pt, label={\scriptsize$\triangleright$}]
        \item \ref{app:loop_scaling}\; Full Loop Scaling Results
        \item \ref{app:loss_landscape}\; Loss Landscape Analysis
        \item \ref{app:loop_diagnostics}\; Per-Loop Diagnostics
    \end{itemize}
    \item \textbf{Appendix~\ref{app:moe_analysis}.\; MoE Analysis}
    \begin{itemize}[leftmargin=1.2em, itemsep=0pt, parsep=0pt, topsep=1pt, label={\scriptsize$\triangleright$}]
        \item \ref{app:balance_loss}\; Load-Balancing Auxiliary Loss
        \item \ref{app:moe_hyper}\; Parameter Sensitivity Analysis
        \item \ref{app:expert_routing}\; Expert Routing Analysis
    \end{itemize}
    \item \textbf{Appendix~\ref{app:hc_analysis}.\; Hyper-Connected Residuals Analysis}
    \item \textbf{Appendix~\ref{app:efficiency_impl}.\; Efficiency \& Implementation}
    \begin{itemize}[leftmargin=1.2em, itemsep=0pt, parsep=0pt, topsep=1pt, label={\scriptsize$\triangleright$}]
        \item \ref{app:impl}\; Implementation Details
        \item \ref{app:complexity}\; Complexity Analysis
        \item \ref{app:efficiency}\; Efficiency Comparison
    \end{itemize}
    \item \textbf{Appendix~\ref{app:related_work}.\; Extended Related Work}
    \begin{itemize}[leftmargin=1.2em, itemsep=0pt, parsep=0pt, topsep=1pt, label={\scriptsize$\triangleright$}]
        \item \ref{app:related_work_ctr}\; Transformer-based CTR Prediction
        \item \ref{app:related_work_loop}\; Looped Transformers
    \end{itemize}
\end{itemize}
\end{tcolorbox}

\section{Loop Scaling Deep Dive}
\label{app:loop_deep_dive}

\subsection{Full Loop Scaling Results}
\label{app:loop_scaling}

Table~\ref{tab:loop_scaling_full} reports the complete loop scaling results across training loop counts $L \in \{0,1,2,3\}$. For models trained with loops ($L>0$), we evaluate inference loop counts $i \in \{0, 1, 2, 3\}$; for $L=0$, only zero-loop inference is applicable.

\begin{table}[h]
\centering
\caption{Full loop scaling results. Each block corresponds to a training loop count $L$. $i$: inference loops. Oracle selects the optimal loop depth per sample.}
\label{tab:loop_scaling_full}
\resizebox{\linewidth}{!}{
\setlength{\tabcolsep}{3.5pt}
\begin{tabular}{cl|ccc|ccc|ccc|ccc}
\toprule
\multirow{2}{*}{$L$} & \multirow{2}{*}{\textbf{Config}} & \multicolumn{3}{c|}{\textbf{Amazon}} & \multicolumn{3}{c|}{\textbf{TaobaoAds}} & \multicolumn{3}{c|}{\textbf{KuaiVideo}} & \multicolumn{3}{c}{\textbf{InHouse}} \\
\cmidrule(lr){3-5} \cmidrule(lr){6-8} \cmidrule(lr){9-11} \cmidrule(lr){12-14}
& & AUC\,$\uparrow$ & GAUC\,$\uparrow$ & NE\,$\downarrow$ & AUC\,$\uparrow$ & GAUC\,$\uparrow$ & NE\,$\downarrow$ & AUC\,$\uparrow$ & GAUC\,$\uparrow$ & NE\,$\downarrow$ & AUC\,$\uparrow$ & GAUC\,$\uparrow$ & NE\,$\downarrow$ \\
\midrule
\rowcolor{L0bg} 0 & $i{=}0$ & .8662 & .8642 & .6682 & .6423 & .5661 & .9701 & .7440 & .6624 & .8791 & .6966 & .5661 & .9221 \\
\midrule
\rowcolor{L1bg}   & $i{=}0$ & .8683 & .8674 & .6644 & .6436 & .5662 & .9693 & .7445 & .6631 & .8784 & .6985 & .5688 & .9211 \\
\rowcolor{L1bg}   & $i{=}1$ & .8687 & .8677 & .6641 & .6433 & .5658 & .9695 & .7447 & .6634 & .8783 & .6984 & .5691 & .9210 \\
\rowcolor{L1bg} 1 & $i{=}2$ & .8683 & .8672 & .6681 & .6425 & .5652 & .9702 & .7445 & .6631 & .8785 & .6982 & .5701 & .9211 \\
\rowcolor{L1bg}   & $i{=}3$ & .8668 & .8660 & .6784 & .6414 & .5647 & .9711 & .7436 & .6619 & .8792 & .6978 & .5707 & .9220 \\
\rowcolor{oraclebg}   & \textit{Oracle} & \textit{.8885} & \textit{.8877} & \textit{.6138} & \textit{.6744} & \textit{.6522} & \textit{.9520} & \textit{.7675} & \textit{.7119} & \textit{.8488} & \textit{.7306} & \textit{.6908} & \textit{.8908} \\
\midrule
\rowcolor{L2bg}   & $i{=}0$ & .8670 & .8660 & .6588 & .6438 & .5659 & .9705 & .7443 & .6630 & .8787 & .7000 & .5696 & .9186 \\
\rowcolor{L2bg}   & $i{=}1$ & .8695 & .8685 & .6539 & .6436 & .5662 & .9707 & .7445 & .6633 & .8785 & .7002 & .5699 & .9185 \\
\rowcolor{L2bg} 2 & $i{=}2$ & .8701 & .8688 & .6527 & .6432 & .5663 & .9710 & .7445 & .6631 & .8785 & .7002 & .5697 & .9185 \\
\rowcolor{L2bg}   & $i{=}3$ & .8697 & .8683 & .6529 & .6428 & .5663 & .9712 & .7443 & .6627 & .8785 & .7000 & .5695 & .9187 \\
\rowcolor{oraclebg}   & \textit{Oracle} & \textit{.8865} & \textit{.8855} & \textit{.6101} & \textit{.6701} & \textit{.6385} & \textit{.9556} & \textit{.7614} & \textit{.6985} & \textit{.8566} & \textit{.7201} & \textit{.6461} & \textit{.9009} \\
\midrule
\rowcolor{L3bg}   & $i{=}0$ & .8715 & .8700 & .6594 & .6439 & .5666 & .9694 & .7448 & .6635 & .8774 & .7007 & .5687 & .9187 \\
\rowcolor{L3bg}   & $i{=}1$ & .8728 & .8713 & .6571 & .6441 & .5664 & .9691 & .7450 & .6640 & .8774 & .7007 & .5691 & .9185 \\
\rowcolor{L3bg} 3 & $i{=}2$ & .8728 & .8715 & .6567 & .6440 & .5664 & .9691 & .7449 & .6639 & .8774 & .7005 & .5690 & .9187 \\
\rowcolor{L3bg}   & $i{=}3$ & .8726 & .8713 & .6560 & .6436 & .5662 & .9693 & .7448 & .6638 & .8774 & .7002 & .5688 & .9190 \\
\rowcolor{oraclebg}   & \textit{Oracle} & \textit{.8858} & \textit{.8844} & \textit{.6207} & \textit{.6672} & \textit{.6303} & \textit{.9569} & \textit{.7588} & \textit{.6942} & \textit{.8591} & \textit{.7195} & \textit{.6414} & \textit{.9013} \\
\bottomrule
\end{tabular}
}
\end{table}

\subsection{Loss Landscape Analysis}
\label{app:loss_landscape}

\begin{figure}[h]
    \centering
    \includegraphics[width=0.95\linewidth]{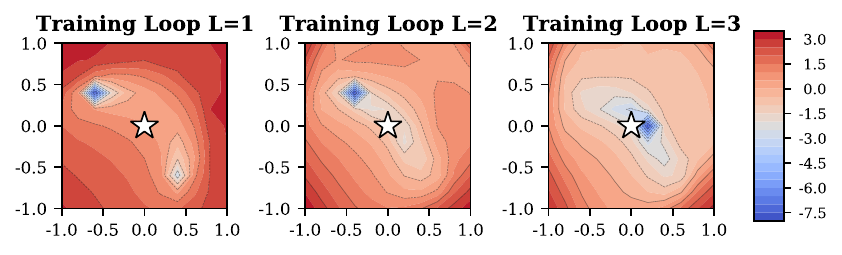}
    \vspace{-0.8em}
    \caption{Loss landscape visualization on Amazon with varying training loop counts ($L{=}1,2,3$) at full inference depth. Warmer colors indicate higher loss; $\star$ marks the converged optimum. More training loops produce broader, flatter minima.}
    \label{fig:loss_landscape}
\end{figure}

Figure~\ref{fig:loss_landscape} visualizes the loss landscape on the Amazon dataset using the filter-normalized random direction method~\citep{li2018visualizing}. We compare models trained with different loop counts ($L{=}1, 2, 3$) at full inference depth. The $L{=}1$ model exhibits the smallest low-loss basin with a secondary local minimum visible in the lower-right region, indicating a more rugged optimization landscape. As $L$ increases, the low-loss basin progressively broadens and the landscape becomes smoother: by $L{=}3$, the blue region around the optimum is substantially wider with more evenly spaced contours. Since broader, flatter minima are generally associated with better generalization~\citep{li2018visualizing,lee2024visualizinglosslandscapeselfsupervised,10.1145/3604915.3608831,na-etal-2022-train,10205050}, this provides a geometric explanation for why more training loops yield higher realized performance (Section~\ref{sec:loop_scaling}).

\subsection{Per-Loop Diagnostics}
\label{app:loop_diagnostics}

\begin{wrapfigure}{r}{0.42\textwidth}
    \vspace{-1.5em}
    \centering
    \includegraphics[width=0.43\textwidth]{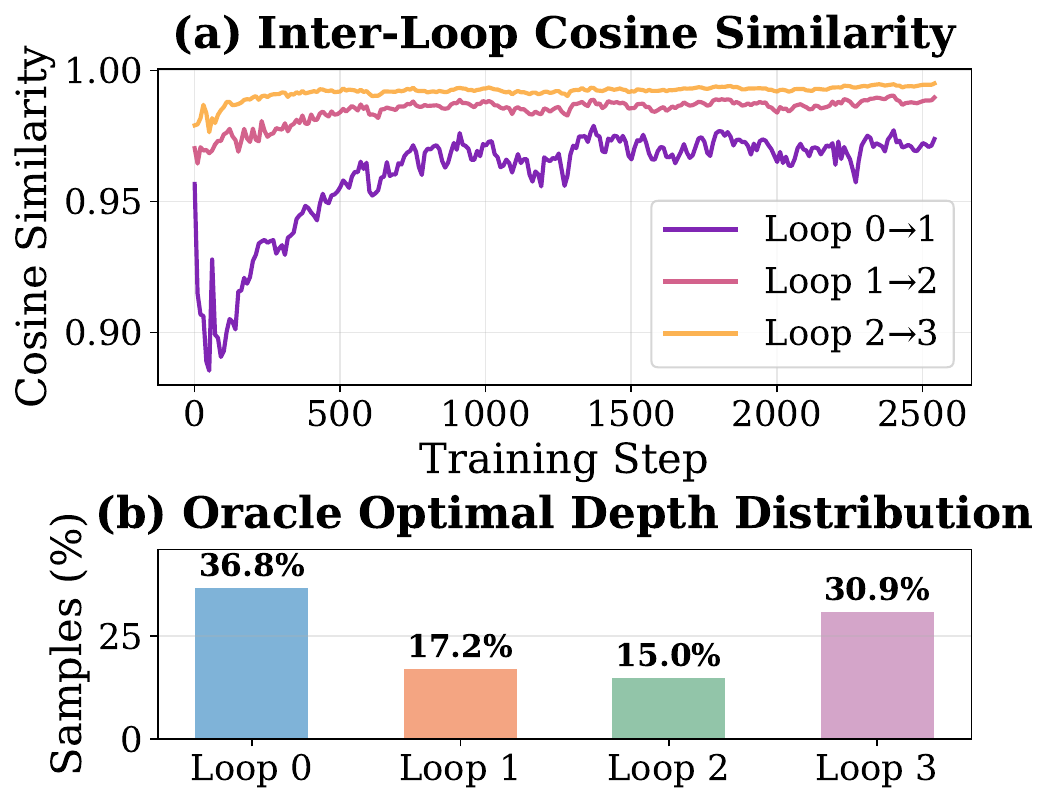}

    \caption{Per-loop diagnostics. \textbf{(a)} Cosine similarity between adjacent loop depths during training. \textbf{(b)} Oracle optimal loop-depth distribution on test set.}
    \label{fig:loop_loss_sim}
    \vspace{-1.em}
\end{wrapfigure}

Figure~\ref{fig:loop_loss_sim} provides two views of how the Loop Block differentiates its behavior across depths.

\paragraph{(a) Inter-loop representation similarity.}
The cosine similarity between adjacent loop depths' global token representations increases during training, indicating that the shared Loop Block \textbf{progressively aligns representations across depths}. However, the similarity does not reach 1.0, meaning that each loop depth retains a distinct representation. The similarity between later loops (loop 2$\rightarrow$3) is higher than between earlier ones (loop 0$\rightarrow$1), suggesting that iterative refinement converges as depth increases, consistent with the diminishing returns observed in inference loop scaling. This progressive alignment can be viewed as a form of \textbf{implicit self-distillation} within the model: deeper loop iterations act as ``teachers'' that guide shallower iterations toward better representations through the shared parameters, which helps explain why even the zero-loop output achieves strong performance.

\paragraph{(b) Oracle optimal depth distribution.}
The oracle analysis reveals a notably \textbf{non-uniform distribution}: 36.8\% of samples achieve their best prediction at loop 0 (zero-loop) and 30.9\% at loop 3 (full depth), while loops 1 and 2 account for 17.2\% and 15.0\%, respectively. This \textbf{bimodal pattern} suggests that the sample population naturally splits into two groups: those that are already well-predicted by the Entry Block alone, and those that benefit from the full iterative refinement. Notably, 36.8\% of samples are best served at loop 0, indicating that for a substantial fraction of inputs, additional loop iterations do not improve and may even degrade predictions. This is analogous to the \textbf{overthinking} phenomenon observed in reasoning models~\citep{tang2025think,sui2025stop}, where excessive computation on already-solved instances wastes resources without quality gains. As shown in Table~\ref{tab:overall}, the oracle upper bound still enjoys a \textbf{substantial margin} over the best realized inference (e.g., 0.8858 vs.\ 0.8728 AUC on Amazon), indicating that a large portion of this headroom remains untapped and could be unlocked by future \textbf{adaptive inference} strategies that selectively allocate loop depth per sample.

\section{MoE Analysis}
\label{app:moe_analysis}

\subsection{Load-Balancing Auxiliary Loss}
\label{app:balance_loss}

In sparse MoE, an unconstrained router may collapse to routing all tokens to a small subset of experts, leaving others underutilized. To prevent this, we adopt the load-balancing auxiliary loss proposed in~\citep{fedus2022switch}. For a batch of $N$ tokens, a router with $E$ experts, and top-$k$ routing, we define:
\begin{itemize}[leftmargin=2em]
    \item $f_e$: the \emph{dispatch-normalized usage} of expert $e$, \textit{i.e.}, the fraction of all top-$k$ routing assignments dispatched to expert $e$:
    \begin{equation}
        f_e = \frac{1}{Nk} \sum_{i=1}^{N} \mathbf{1}[e \in \text{top-}k(\mathbf{r}_i)],
    \end{equation}
    where $\mathbf{r}_i$ is the router logit vector for token $i$.
    \item $p_e$: the \emph{mean router probability} for expert $e$, computed by averaging the softmax-normalized router probabilities across all tokens:
    \begin{equation}
        p_e = \frac{1}{N} \sum_{i=1}^{N} \text{softmax}(\mathbf{r}_i)_e.
    \end{equation}
\end{itemize}
The load-balancing loss is then defined as:
\begin{equation}
    \mathcal{L}_{\text{bal}} = E \cdot \sum_{e=1}^{E} f_e \cdot p_e.
\end{equation}
When all experts are utilized equally, $f_e = 1/E$ and $p_e = 1/E$ for all $e$, yielding $\mathcal{L}_{\text{bal}} = 1$. Any deviation from uniform routing increases this loss, thereby encouraging balanced expert utilization. The final training objective combines the multi-depth BCE loss with the load-balancing term:
\begin{equation}
    \mathcal{L} = \mathcal{L}_{\text{total}} + \lambda \cdot \mathcal{L}_{\text{bal}},
\end{equation}
where $\lambda$ is a hyperparameter that controls the strength of the load-balancing regularization.

\subsection{Parameter Sensitivity Analysis}
\label{app:moe_hyper}

We study the sensitivity of LoopCTR to two key hyperparameters: the number of activated experts (with total experts fixed at 4) and the total number of experts (with activated experts fixed at 2). Table~\ref{tab:moe_hyper} reports the results on Amazon and KuaiVideo.

\begin{table}[h]
\centering
\caption{Parameter sensitivity analysis on Amazon and KuaiVideo. \textbf{Top}: varying activated experts with 4 total experts. \textbf{Bottom}: varying total experts with 2 activated experts. The notation $k$/$n$ denotes $k$ activated out of $n$ total experts.}
\label{tab:moe_hyper}
\setlength{\tabcolsep}{5pt}
\begin{tabular}{l|ccc|ccc}
\toprule
\multirow{2}{*}{\textbf{Config}} & \multicolumn{3}{c|}{\textbf{Amazon}} & \multicolumn{3}{c}{\textbf{KuaiVideo}} \\
\cmidrule(lr){2-4} \cmidrule(lr){5-7}
& AUC\,$\uparrow$ & GAUC\,$\uparrow$ & NE\,$\downarrow$ & AUC\,$\uparrow$ & GAUC\,$\uparrow$ & NE\,$\downarrow$ \\
\midrule
\multicolumn{7}{l}{\textit{Varying activated experts (total = 4)}} \\
\midrule
1 / 4 & .8660 & .8648 & .6678 & .7440 & .6616 & .8795 \\
2 / 4 & \textbf{.8726} & \textbf{.8713} & \textbf{.6560} & \textbf{.7448} & \textbf{.6638} & \textbf{.8774} \\
3 / 4 & .8690 & .8678 & .6631 & .7428 & .6624 & .8810 \\
4 / 4 & .8670 & .8655 & .6716 & .7437 & .6637 & .8792 \\
\midrule
\multicolumn{7}{l}{\textit{Varying total experts (activated = 2)}} \\
\midrule
2 / 2 & .8719 & \textbf{.8713} & \textbf{.6492} & .7441 & .6600 & .8816 \\
2 / 3 & .8705 & .8682 & .6576 & .7423 & .6593 & .8833 \\
2 / 4 & \textbf{.8726} & \textbf{.8713} & .6560 & \textbf{.7448} & \textbf{.6638} & .8774 \\
2 / 5 & .8688 & .8674 & .6607 & .7446 & .6624 & \textbf{.8759} \\
\bottomrule
\end{tabular}
\end{table}

\paragraph{Activated experts.} Activating 2 out of 4 experts yields the best AUC on both datasets. Using only 1 expert (no routing diversity) and activating all 4 experts (no sparsity) both degrade performance, confirming the importance of sparse expert selection for balancing capacity and regularization.

\paragraph{Total experts.} With 2 activated experts, increasing the total from 2 to 4 improves performance (Amazon AUC: 0.8719 $\rightarrow$ 0.8726; KuaiVideo AUC: 0.7441 $\rightarrow$ 0.7448), but further increasing to 5 offers no additional gain. The 2/4 configuration provides the best trade-off between expert diversity and training stability.

\subsection{Expert Routing Analysis}
\label{app:expert_routing}

\begin{figure}[h]
    \centering
    \includegraphics[width=0.9\linewidth]{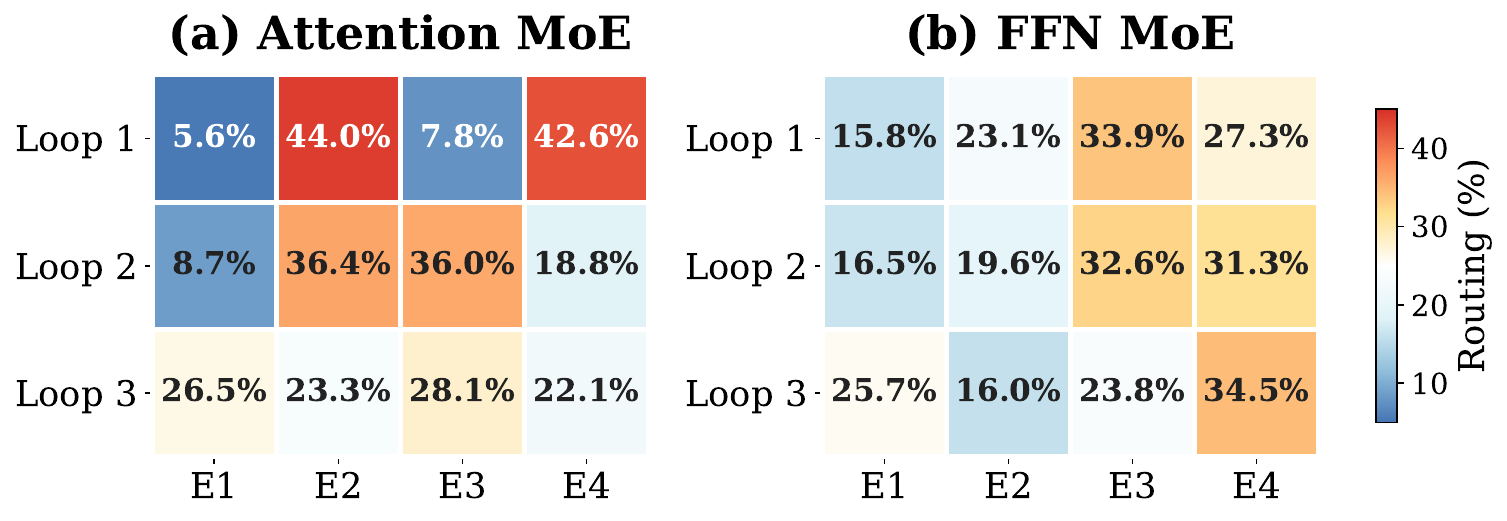}
    \vspace{-0.8em}
    \caption{Expert routing distribution across loop iterations on Amazon with $L{=}3$. \textbf{Left}: attention MoE. \textbf{Right}: FFN MoE. Each bar shows the dispatch-normalized percentage of top-$k$ routing assignments received by each of the 4 experts.}
    \label{fig:expert_routing}
\end{figure}

To understand how the shared MoE layer adapts its routing across loop iterations, we visualize the expert activation distribution for both attention and FFN MoE in LoopCTR($L{=}3$) on Amazon. Figure~\ref{fig:expert_routing} shows the dispatch-normalized percentage of top-$k$ routing assignments received by each expert at each loop iteration.

\paragraph{Attention MoE.} The routing distribution shifts substantially across iterations. At iteration 1, routing is highly concentrated on experts 2 and 4 (44.0\% and 42.6\%), while experts 1 and 3 are rarely activated. By iteration 3, the distribution becomes nearly uniform ($\sim$22--28\% per expert). This suggests that early iterations rely on specialized expert pathways, while later iterations require more balanced computation as representations become increasingly refined.

\paragraph{FFN MoE.} The FFN routing is comparatively more balanced from the start ($\sim$16--34\% per expert), though it still evolves across iterations. Expert 3 dominates in iterations 1--2 (33.9\% and 32.6\%) but its share decreases in iteration 3, while expert 4's share grows. This gradual shift indicates that the shared FFN layer dynamically adjusts its computational pathways at different loop depths, further confirming that the Loop Block does not simply repeat identical computation but adapts its behavior across different iteration depths.

\section{Hyper-Connected Residuals Analysis}
\label{app:hc_analysis}

\begin{figure}[h]
    \centering
    \includegraphics[width=\linewidth]{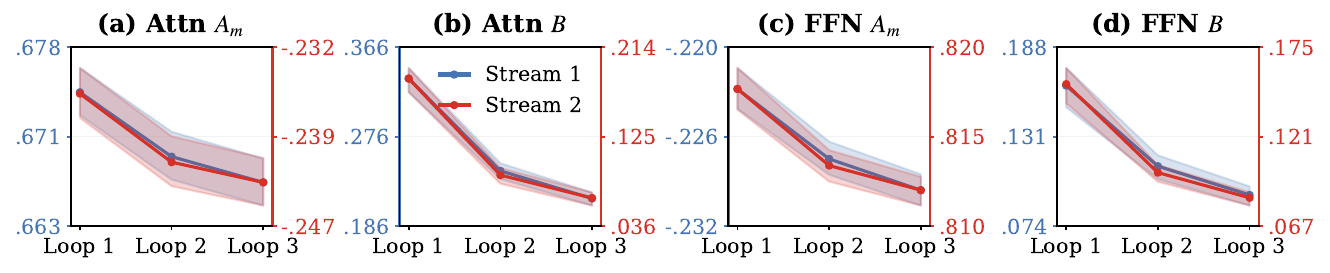}
    \vspace{-1.5em}
    \caption{Visualization of the learned residual-stream coefficients in HCR for the Attention and FFN sub-layers. Lines denote the mean and shaded regions indicate the 30th--70th percentile range.}
    \label{fig:hc}
\end{figure}

A core design choice in LoopCTR is replacing the standard residual $\mathbf{h} + f(\mathbf{h})$ with Hyper-Connected Residuals (HCR), which provide multi-stream adaptive residual connections with input-dependent coefficients controlling how each stream flows through the attention and FFN sub-layers (Figure~\ref{fig:architecture}). This brings three key benefits for looped architectures:
(1)~\textbf{Parallel multi-stream computation.} The $n$ parallel streams improve hardware utilization over the single-stream residual.
(2)~\textbf{Flexible blending.} The mixing matrix $\mathbf{A}_r \in \mathbb{R}^{n \times n}$ replaces the fixed 1:1 skip ratio with learned, per-stream blending coefficients.
(3)~\textbf{Input-dependent adaptivity.} The input-dependent coefficients make the residual {\em instance-adaptive} and implicitly {\em loop-aware}, allowing the shared layer to modulate information flow differently at each iteration without explicit loop index conditioning.
As visualized in Figure~\ref{fig:hc}, the learned coefficients exhibit distinct distributions across the attention and FFN sub-layers and vary across loop iterations, confirming that HCR enables the shared layer to differentiate its residual behavior at each loop depth.

\section{Efficiency \& Implementation}
\label{app:efficiency_impl}

\subsection{Implementation Details}
\label{app:impl}

All experiments are conducted on 8 NVIDIA H20 GPUs. We use AdamW~\citep{loshchilov2017decoupled} as the optimizer with a fixed learning rate of 0.001 and batch size of 2048. The embedding dimension is set to 64. For MoE, the number of total experts is searched in $\{2, 3, 4, 5\}$, and an auxiliary load-balancing loss is applied with a regularization weight tuned from $\{0, 0.001, 0.01, 0.1\}$. Hyper-Connected Residuals use $n{=}2$ streams, one for the attention sub-layer and one for the FFN sub-layer, enabling shared-parameter loop inference. On the InHouse dataset, the number of learnable query tokens for long-term sequence compression is fixed at 16.

\subsection{Complexity Analysis}
\label{app:complexity}

We analyze the computational complexity of LoopCTR. Let $T_{\text{seq}}$ and $T_{\text{glb}}$ denote the number of sequential and global tokens after long-term sequence compression, respectively, and let $T=T_{\text{seq}}+T_{\text{glb}}$. Let $d$ denote the hidden dimension, $d_{\text{ff}}$ the FFN intermediate dimension, $n$ the number of hyper-connection streams, $E$ the total number of MoE experts, $k$ the number of activated experts per token, $L$ the number of training loops, and $i$ the number of inference loops. In production serving, user-side computations such as long-term sequence compression and Entry Block processing of behavior sequences can be amortized or cached across candidate items within the same request, while item-side, context, and user-item cross features are computed online.\footnote{A further engineering direction is to decouple user-side and item-side feature processing more aggressively, exposing more user-only computation to offline precomputation or request-level caching and thereby reducing online cost and latency.}

\begin{table}[h]
\centering
\caption{Per-component complexity of LoopCTR. $m_j$: tokens in Entry Block group $j$; $k$: activated experts; $n$: HCR streams. In online serving, user-side computations ($C_{\text{user}}$, shaded \colorbox{L0bg}{gray}) are cached once per request and shared across $N$ candidate items; only per-item computations ($C_{\text{item}}$, unshaded) are executed for each item.}
\label{tab:complexity_summary}
\resizebox{\linewidth}{!}{
\setlength{\tabcolsep}{5pt}
\begin{tabular}{l l l l}
\toprule
\textbf{Component} & \textbf{Sub-computation} & \textbf{Complexity} & \textbf{Online} \\
\midrule
\multirow{3}{*}{Entry Block}
& \cellcolor{L0bg} Seq.\ group projection + self-attention & \cellcolor{L0bg} $O\big(\sum_j m_j^2 d + (1{+}k) T_{\text{seq}} d^2 + k T_{\text{seq}} d\, d_{\text{ff}}\big)$ & \cellcolor{L0bg} $C_{\text{user}}$ \\
\cmidrule(l){2-4}
& Global token projection + FFN & $O\big((1{+}k) T_{\text{glb}} d^2 + k T_{\text{glb}} d\, d_{\text{ff}}\big)$ & $C_{\text{item}}$ \\
\midrule
\multirow{3}{*}{\shortstack[l]{Loop Block\\($\times i$ iters)}}
& \cellcolor{L0bg} Seq-to-seq attention + seq FFN/HCR & \cellcolor{L0bg} $O\big(T_{\text{seq}}^2 d + (1{+}k) T_{\text{seq}} d^2 + k T_{\text{seq}} d\, d_{\text{ff}} + T_{\text{seq}} n^2 d\big)$ & \cellcolor{L0bg} $C_{\text{user}}$ \\
\cmidrule(l){2-4}
& Global-to-all attention + global FFN/HCR & $O\big(T_{\text{glb}} T d + (1{+}k) T_{\text{glb}} d^2 + k T_{\text{glb}} d\, d_{\text{ff}} + T_{\text{glb}} n^2 d\big)$ & $C_{\text{item}}$ \\
\midrule
Exit Block
& Cross-attn + FFN + tower (all per-item) & $O\big(T_{\text{glb}} T_{\text{seq}} d + (1{+}k)(T_{\text{glb}}{+}T_{\text{seq}}) d^2 + k T_{\text{glb}} d\, d_{\text{ff}} + C_{\text{tower}}\big)$ & $C_{\text{item}}$ \\
\midrule
\multicolumn{4}{l}{\textbf{Total} ($i$ inference loops)} \\
\quad Full forward & \multicolumn{3}{l}{$C_{\text{entry}} + i \cdot C_{\text{loop}} + C_{\text{exit}}$; \; zero-loop ($i{=}0$): $C_{\text{entry}} + C_{\text{exit}}$} \\
\quad Online serving & \multicolumn{3}{l}{$N \cdot C_{\text{item}}$ \; ($C_{\text{user}}$ cached once per request and shared across $N$ items)} \\
\bottomrule
\end{tabular}
}
\end{table}

\paragraph{Entry Block.}
The Entry Block applies heterogeneous feature projections, grouped self-attention, and an FFN, with MoE used in the value/output projections and FFN. Since each group attends independently, the attention-score cost is $O(\sum_j m_j^2 d)$, where $m_j$ is the number of tokens in group $j$. Heterogeneous feature projections and dense query/key projections contribute $O(Td^2)$, sparse MoE value/output projections contribute $O(kTd^2)$, and the FFN MoE contributes $O(kT d d_{\text{ff}})$. HCR adds $O(Tn^2d)$ across the attention and FFN sub-layers. Thus,
\begin{equation}
    C_{\text{entry}} =
    O\!\left(\sum_j m_j^2 d + T d^2 + kT d^2 + kT d d_{\text{ff}} + Tn^2d\right).
\end{equation}
Because groups are processed independently and each group is small (the short-term sequence, compressed long-term query tokens, or individual global tokens), the attention-score cost is typically much less than full-sequence attention.

\paragraph{Loop Block.}
Each loop iteration applies prefix attention and an FFN, both augmented with Hyper-Connected Residuals and MoE. Sequential tokens attend among themselves, while global tokens attend to both sequential and global tokens. The attention-score cost is therefore $O(T_{\text{seq}}^2 d + T_{\text{glb}}Td)$. Dense query/key projections contribute $O(Td^2)$, and $k$-out-of-$E$ MoE value/output projections contribute $O(kTd^2)$ rather than $O(ETd^2)$.

\textit{FFN MoE.} The FFN uses the same $k$-out-of-$E$ sparse routing, giving a per-token cost of $O(kd \cdot d_{\text{ff}})$ where $d_{\text{ff}}$ is the FFN intermediate dimension, compared to $O(Ed \cdot d_{\text{ff}})$ if all experts were activated.

\textit{Hyper-Connected Residuals.} Computing the dynamic coefficients and multi-stream residual mixing costs $O(Tn^2d)$ per sub-layer, dominated by the dynamic residual-mixing projection and stream mixing. With $n{=}2$ in our experiments, this overhead is negligible compared to attention and FFN.

Combining these terms, the per-iteration Loop Block cost is
\begin{equation}
    C_{\text{loop}} =
    O\!\left(
    T_{\text{seq}}^2 d
    + T_{\text{glb}}T d
    + T d^2
    + kT d^2
    + kT d d_{\text{ff}}
    + Tn^2d
    \right).
\end{equation}

\paragraph{Exit Block.}
The Exit Block applies cross-attention from global queries to sequential keys/values, followed by an FFN sub-layer and a small task tower. Its attention-score cost is $O(T_{\text{glb}}T_{\text{seq}}d)$. Dense query and key projections are applied to global and sequential tokens, respectively, giving $O((T_{\text{glb}}+T_{\text{seq}})d^2)$. The MoE value projection is applied to sequential tokens ($O(kT_{\text{seq}}d^2)$), while the MoE output projection is applied to the global cross-attention outputs ($O(kT_{\text{glb}}d^2)$). The FFN MoE operates on the global outputs and costs $O(kT_{\text{glb}}d d_{\text{ff}})$. Therefore,
\begin{equation}
    C_{\text{exit}} =
    O\!\left(
    T_{\text{glb}}T_{\text{seq}}d
    + (T_{\text{glb}}+T_{\text{seq}})d^2
    + kT_{\text{seq}}d^2
    + kT_{\text{glb}}d^2
    + kT_{\text{glb}}d d_{\text{ff}}
    + C_{\text{tower}}
    \right),
\end{equation}
where $C_{\text{tower}}$ denotes the cost of the final task tower.

\paragraph{Overall cost: full forward pass.}
At training time (or offline evaluation), the Entry Block is executed once, the Loop Block is executed $L$ times, and the Exit Block is invoked $L{+}1$ times (once per loop depth for process supervision). The total cost is therefore $C_{\text{entry}} + L \cdot C_{\text{loop}} + (L{+}1) \cdot C_{\text{exit}}$, scaling linearly with $L$. At inference time with $i$ loops, the cost is $C_{\text{entry}} + i \cdot C_{\text{loop}} + C_{\text{exit}}$; in the zero-loop mode ($i{=}0$), the Loop Block is bypassed entirely, reducing the cost to $C_{\text{entry}} + C_{\text{exit}}$.

\paragraph{Overall cost: online serving with KV caching.}
In production serving, the asymmetric attention mask in the Loop Block (sequential tokens never attend to global tokens) enables a key optimization: the sequential KV states can be computed once per user request and shared across all candidate items. Concretely, the Entry Block processing of behavior sequences and the sequential-to-sequential attention in the Loop Block are user-side computations that are independent of item features. These can be computed once and cached, so that scoring each candidate item only requires the item-side global token computations. Let $C_{\text{user}}$ and $C_{\text{item}}$ denote the user-side and per-item costs, respectively. For a request with $N$ candidate items, the full forward cost is $N \cdot (C_{\text{user}} + C_{\text{item}})$, while with caching the cost reduces to $C_{\text{user}} + N \cdot C_{\text{item}}$. Since $T_{\text{seq}} \gg T_{\text{glb}}$ in practice and the sequential attention dominates the computation, this caching strategy yields substantial savings: the amortized per-item cost is $C_{\text{item}} + C_{\text{user}} / N \approx C_{\text{item}}$ for large $N$.

\paragraph{Additional parameters from LoopCTR components.}
Compared to a standard Transformer layer, LoopCTR introduces two sources of additional parameters:
\begin{itemize}[leftmargin=2em]
    \item \textbf{Hyper-Connected Residuals}: for each sub-layer, static parameters $\mathbf{A}_m$ ($n{\times}1$), $\mathbf{A}_r$ ($n{\times}n$), $\mathbf{B}$ ($1{\times}n$), and dynamic projections $\mathbf{W}_m$ ($d{\times}1$), $\mathbf{W}_r$ ($d{\times}n$), $\mathbf{W}_{\beta}$ ($d{\times}1$), plus scalars $s_{\alpha}$, $s_{\beta}$. With $n{=}2$, each sub-layer adds $4d + 10$ parameters, which is negligible compared to the $\mathcal{O}(d^2)$ parameters of the sub-layer itself.
    \item \textbf{MoE}: expanding a projection from a single matrix to $E$ expert matrices multiplies its parameter count by $E$, while each token only activates $k$ experts during computation due to sparse routing. For attention MoE (V and O projections) and FFN MoE, the additional parameters are $(E{-}1) \cdot (2d^2 + 2d \cdot d_{\text{ff}})$.
\end{itemize}
Crucially, all Loop Block parameters are shared across loop iterations, so the full model parameter count is \emph{independent of $L$} and of the nonzero inference loop count. This stands in contrast to existing CTR scaling approaches that increase model depth by stacking heterogeneous layers with distinct parameters, where $L$ layers require $L$ times the per-layer parameter budget. For strict zero-loop deployment, the Loop Block is never executed and can be omitted from the deployed subnetwork; Table~\ref{tab:efficiency} therefore reports active/deployed parameters, which are smaller for LoopCTR(0/3).

\subsection{Efficiency Comparison}
\label{app:efficiency}

\begin{table}[t]
\centering
\caption{Efficiency comparison. Params: active/deployed dense parameter count (M); FLOPs: per-sample floating-point operations (M); Latency: inference time per batch of 2048 samples (ms). For Oracle, FLOPs and Latency are computed as weighted averages based on the per-sample optimal loop depth distribution.}
\label{tab:efficiency}
\resizebox{\linewidth}{!}{
\setlength{\tabcolsep}{3pt}
\begin{tabular}{l|ccc|ccc|ccc|ccc}
\toprule
\multirow{2}{*}{\textbf{Method}} & \multicolumn{3}{c|}{\textbf{Amazon}} & \multicolumn{3}{c|}{\textbf{TaobaoAds}} & \multicolumn{3}{c|}{\textbf{KuaiVideo}} & \multicolumn{3}{c}{\textbf{InHouse}} \\
\cmidrule(lr){2-4} \cmidrule(lr){5-7} \cmidrule(lr){8-10} \cmidrule(lr){11-13}
& Params & FLOPs & Lat. & Params & FLOPs & Lat. & Params & FLOPs & Lat. & Params & FLOPs & Lat. \\
\midrule
DLRM        & 0.99 & 1.97 & 0.89  & 1.91 & 3.80 & 1.12   & 1.12 & 3.88 & 1.81   & 3.22 & 6.42 & 8.46 \\
DIN         & 1.02 & 8.94 & 2.45  & 1.94 & 7.29 & 1.93   & 1.19 & 17.83 & 4.99  & 0.26 & 78.45 & 24.91 \\
DCNv2       & 0.99 & 2.38 & 1.04  & 1.91 & 12.68 & 2.06  & 1.12 & 5.08 & 2.00   & 0.16 & 37.70 & 11.27 \\
Wukong      & 1.50 & 3.29 & 1.66  & 2.54 & 5.77 & 2.23   & 2.29 & 6.75 & 2.98   & 2.95 & 6.87 & 9.47 \\
\midrule
DHEN        & 0.69 & 2.56 & 2.12  & 3.15 & 9.18 & 2.97   & 1.09 & 5.68 & 3.30   & 25.47 & 63.28 & 19.56 \\
AutoInt     & 0.59 & 5.98 & 1.78  & 0.60 & 23.53 & 3.87  & 0.60 & 10.06 & 2.91  & 0.60 & 50.70 & 13.91 \\
HiFormer    & 0.72 & 1.64 & 1.56  & 2.38 & 5.61 & 2.44   & 0.96 & 3.84 & 2.42   & 4.74 & 11.63 & 11.27 \\
\midrule
OneTrans    & 1.05 & 20.88 & 9.81 & 1.01 & 10.96 & 5.13  & 0.81 & 38.27 & 16.85 & 1.29 & 417.97 & 494.58 \\
HSTU        & 0.15 & 42.75 & 15.77 & 0.15 & 22.19 & 8.13 & 0.15 & 117.31 & 43.27 & 0.15 & 2150.00 & 775.72 \\
InterFormer & 0.51 & 7.67 & 8.03  & 2.17 & 7.66 & 4.77   & 0.56 & 17.51 & 9.82  & 5.03 & 162.03 & 479.76 \\
MTGR        & 0.11 & 25.40 & 10.77 & 0.13 & 13.70 & 5.34 & 0.12 & 62.27 & 29.17 & 0.24 & 803.31 & 507.08 \\
\midrule
StackCTR(3) & 1.95 & 201.55 & 29.24 & 1.22 & 69.98 & 51.15 & 1.96 & 414.45 & 33.03 & 2.37 & 18.76 & 120.27 \\
\rowcolor{L3bg}
LoopCTR(0/3) & 0.73 & 62.61 & 6.58 & 0.59 & 22.04 & 3.87 & 0.74 & 124.32 & 10.50 & 1.15 & 13.38 & 9.26 \\
\rowcolor{L3bg}
LoopCTR(1/3) & 1.27 & 108.92 & 14.69 & 0.80 & 38.02 & 29.06 & 1.14 & 221.03 & 19.75 & 1.56 & 15.17 & 73.65 \\
\rowcolor{L3bg}
LoopCTR(2/3) & 1.27 & 155.24 & 21.99 & 0.80 & 54.00 & 40.03 & 1.14 & 317.74 & 26.51 & 1.56 & 16.97 & 97.15 \\
\rowcolor{L3bg}
LoopCTR(3/3) & 1.27 & 201.55 & 29.24 & 0.80 & 69.98 & 50.60 & 1.14 & 414.45 & 33.03 & 1.56 & 18.76 & 120.27 \\
\midrule
\rowcolor{oraclebg}
\textit{Oracle} & \textit{1.27} & \textit{127.34} & \textit{17.28} & \textit{0.80} & \textit{46.54} & \textit{31.15} & \textit{1.14} & \textit{258.65} & \textit{21.22} & \textit{1.56} & \textit{15.94} & \textit{68.95} \\
\bottomrule
\end{tabular}
}
\end{table}

Table~\ref{tab:efficiency} reports the parameter count, FLOPs, and inference latency for all methods. Several observations are worth highlighting.

\paragraph{LoopCTR(0/3) achieves the best efficiency-effectiveness trade-off.}
By bypassing the Loop Block entirely at inference, LoopCTR(0/3) achieves the lowest latency among all LoopCTR variants while already surpassing every baseline on prediction quality (Table~\ref{tab:overall}). On InHouse, LoopCTR(0/3) requires only 13.38M FLOPs and 9.26ms latency, which is $160\times$ fewer FLOPs and $84\times$ lower latency than HSTU (2150M / 775.72ms), and $31\times$ fewer FLOPs and $53\times$ lower latency than OneTrans (417.97M / 494.58ms). This confirms the practical viability of the train-multi-loop, infer-zero-loop.

\paragraph{Parameter sharing reduces active model size.}
Since all executed loop iterations reuse the same Loop Block, LoopCTR's active parameter count remains constant for nonzero inference loops. LoopCTR(3/3) uses the same 1.27M active parameters on Amazon as LoopCTR(1/3), whereas StackCTR(3) requires 1.95M parameters ($1.5\times$ more) for the same FLOPs budget. In strict zero-loop deployment, the Loop Block is bypassed and can be omitted from the deployed subnetwork, which explains the smaller active parameter count of LoopCTR(0/3). This advantage is particularly relevant for deployment scenarios with limited GPU memory.

\paragraph{FLOPs and latency scale linearly with inference loops.}
Each additional inference loop adds a fixed per-loop cost, resulting in a linear relationship between the number of inference loops and both FLOPs and latency. This predictable scaling behavior simplifies resource planning for deployment configurations that trade off latency for accuracy.

\section{Extended Related Work}
\label{app:related_work}

\subsection{Transformer-based CTR Prediction}
\label{app:related_work_ctr}

CTR prediction has evolved through three phases: feature interaction modeling with DNN-based methods~\citep{zhou2018deep,zhou2019deep,wang2021dcn,mao2023finalmlp} and self-attention approaches~\citep{song2019autoint,zhang2021multi}; sequential user behavior modeling via Transformer encoders~\citep{chen2019behavior,zhai2024actions,han2025mtgr,xu2025climber}; and most recently, hybrid architectures that jointly capture feature interactions and sequential patterns~\citep{gui2023hiformer,yu2025hhft,huang2026hyformer,huang2026mixformer,zhang2025onetrans,10.1145/3746252.3761527}. In parallel, industrial scaling efforts such as RankMixer~\citep{zhu2025rankmixer}, Zenith~\citep{zhang2026zenith}, and TokenMixer-Large~\citep{jiang2026tokenmixer} have pushed Transformer-based ranking models to billion-scale parameters. Despite differing in architectural details, all these approaches follow the same scaling philosophy: more parameters and more computation yield better performance, with the two scaling in lockstep. LoopCTR challenges this paradigm by achieving deeper computation through recursive reuse of shared layers rather than parameter accumulation.

\subsection{Looped Transformers}
\label{app:related_work_loop}

The Universal Transformer~\citep{dehghani2018universal} first introduced weight-shared recursive refinement with Adaptive Computation Time (ACT)~\citep{graves2016adaptive} for per-token dynamic halting. \citet{giannou2023looped} later proved that looped Transformers can simulate programmable computers, establishing Turing completeness, and \citet{fan2024looped} demonstrated improved length generalization on algorithmic reasoning tasks. On the practical side, MoEUT~\citep{csordas2024moeut}, LoopLM~\citep{zhu2025scaling}, and ETD~\citep{koishekenov2025encode} have explored training strategies for weight-tied models in language modeling. These works collectively validate the potential of looped architectures, yet they all require executing multiple loops at inference time, and the resulting latency overhead remains largely unaddressed. Moreover, these efforts focus on NLP tasks. To our knowledge, LoopCTR is the first to bring the loop scaling paradigm into the CTR prediction domain, and its process supervision enables a train-multi-loop, infer-zero-loop strategy that resolves the inference cost problem.

\end{document}